\documentclass[%
twocolumn, prb, aps, superscriptaddress, longbibliography,showpacs,amsmath,amssymb,floatfix
]{revtex4-2}

\usepackage{amsfonts}
\usepackage{bbm}
\usepackage{subfigure}
\usepackage{tipa}
\usepackage{color}
\usepackage{epstopdf}
\usepackage{epsfig}
\usepackage{subfigure}
\usepackage{mathrsfs} 
\usepackage{amsmath,amssymb,amsfonts}
\usepackage{graphicx}
\usepackage{dcolumn}
\usepackage{bm}
\usepackage{braket}
\usepackage[colorlinks,linkcolor=blue,anchorcolor=blue,citecolor=blue,urlcolor=blue]{hyperref}

\begin{document}

\title{Spontaneous formation of subsystem and bath under accordion-type driving}
\author{Suyang Lin}
\affiliation{Department of Physics,  University of Science and Technology of China, Hefei 230026, China}
\author{Ming Gong}
\email{gongm@ustc.edu.cn}
\affiliation{Key Laboratory of Quantum Information, University of Science and Technology of China, Hefei 230026, China}
\affiliation{Anhui Province Key Laboratory of Quantum Network, University of Science and Technology of China, Hefei 230026, China}
\affiliation{Hefei National Laboratory, University of Science and Technology of China, Hefei 230088, China}
\affiliation{Synergetic Innovation Center of Quantum Information and Quantum Physics, University of Science and Technology of China, Hefei 230026, China}

\author{Congjun Wu}
\email{wucongjun@westlake.edu.cn}
\affiliation{New Cornerstone Science Laboratory, Department of Physics, School of Science, Westlake University, Hangzhou 310024, Zhejiang, China}
\affiliation{Institute for Theoretical Sciences, Westlake University, Hangzhou 310024, Zhejiang, China}
\affiliation{Key Laboratory for Quantum Materials of Zhejiang Province, School of Science, Westlake University, Hangzhou 310024, Zhejiang, China}
\affiliation{Institute of Natural Sciences, Westlake Institute for Advanced Study, Hangzhou 310024, Zhejiang, China}

\begin{abstract}
Floquet modulations often yield effective Hamiltonians not easily accessible in traditional time-dependent systems, which brings opportunities for exploring novel physics of quantum dynamics. 
We investigate a Floquet system exhibiting translational symmetry at 
any fixed time but the spatial periodicity is time-dependent.
Such a system is a natural platform for studying thermalization and novel dynamical structures.
We find that the single-particle Hilbert space spontaneously develops a structure of a two-level subsystem and the rest part forms a bath.
The dynamic process is analyzed perturbatively within the two-level subsystem as well as numerical solutions, exhibiting stable time-evolutions. 
These results enrich our understanding of Floquet thermalization without definite spatial periodicity, which brings hints for exploring many-body physics such as scar states.  
\end{abstract}
\maketitle 

Periodically driven quantum systems are of fundamental importance for studying non-equilibrium dynamics.
There has appeared tremendous progress along this direction in both condensed matter research
\cite{observationoffloquetstatesingraphene,floquetengineeringofquantummaterials,condensedmatterexsurvivalofFBstatesinthepresenceofscattering} and that of ultra-cold atom \cite{experimentalprethermalfloquet,dotti2024measuringlocalizationphasediagram},
including Floquet engineering and thermalization. 
Floquet engineering \cite{Bukov_review,floquethierarchicalsymmetry,atomicquantumgasesinperiodicallydrivenopticallattices31} provides new methods in realizing celebrated model systems, including the Haldane and Harper ones, which are difficult to achieve in equilibrium states in solids   \cite{experimentalrealizationofhaldanemodel,realizingtheharperhamiltonianinopticallattices,chiralcurrents,photonicfloquettopologicalinsulators,chiralspinliquidphase,tunablegaugepotential}. 
Floquete systems typically thermalize to infinite temperatures \cite{Kuwahara_2016floquetmagnustheory,longtimebehaviorofisolatedperiodicallydriveninteractinglatticesystems}
since energy does not conserve when Hamiltonians are time-dependent \cite{Uedathermalization,statisticalmechanicsoffloquetquantummatterexactandemergentconservationlaws,manybodyonsetoffloquetthermalization}.
Nevertheless, opportunities for 
Floquet systems staying in non-equilibrium steady states still exist, and they escape from thermalization \cite{periodicsteadyregimeandinterferenceinaperiodicallydrivenquantumsystem,absenceofthermalizationinfiniteisolatedinteractingfloquetsystems1,parametricinstabilityinperiodicallydrivenluttingerliquids}, such as 
in systems of time crystals \cite{timecrystal,khemani2019briefhistorytimecrystals}. 
In addition, certain Floquet systems can behave prethermalization, which maintain a characteristic dynamic structure exhibiting a long effective time \cite{na2024engineeringmicromotionfloquetprethermalization,experimentalprethermalfloquet}. 
These observations provide a new possibility of engineering stable quantum states with controllable dynamics.

The symmetry principle is essential for controlling dynamical processes
\cite{floquethierarchicalsymmetry}.
In all situations mentioned above, 
a certain lattice translation symmetry is maintained throughout time modulation.
Partly motivated by the space-time coupled symmetry in Ref. \cite{spacetimecrystal}, we consider a system that at any moment it possesses a lattice symmetry, but the periodicity varies with time.
Such a modulation is termed accordion-type modulation throughout this article, in order to distinguish it from the conventional Floquet one.
In the new situation, the lattice momentum $k$ is no longer well-defined, which would significantly change the thermalization process.

In this article, we investigate the novel effect to the thermalization process brought by the accordion-type modulation.
We focus on the thermalization process and the dynamical structure with the specific symmetry presented in Eq. (\ref{eq-general}).
The system spontaneously develops a stable subspace consisting of two levels under the accordion-type driving.
The quantum dynamics can be understood by employing the perturbative time-dependent eigen wavefunctions.
These findings greatly enrich our understanding the thermalization process, which are helpful for designing new dynamical structures. 

\begin{figure}[htbp]
\centering
\includegraphics[width=0.9\linewidth]{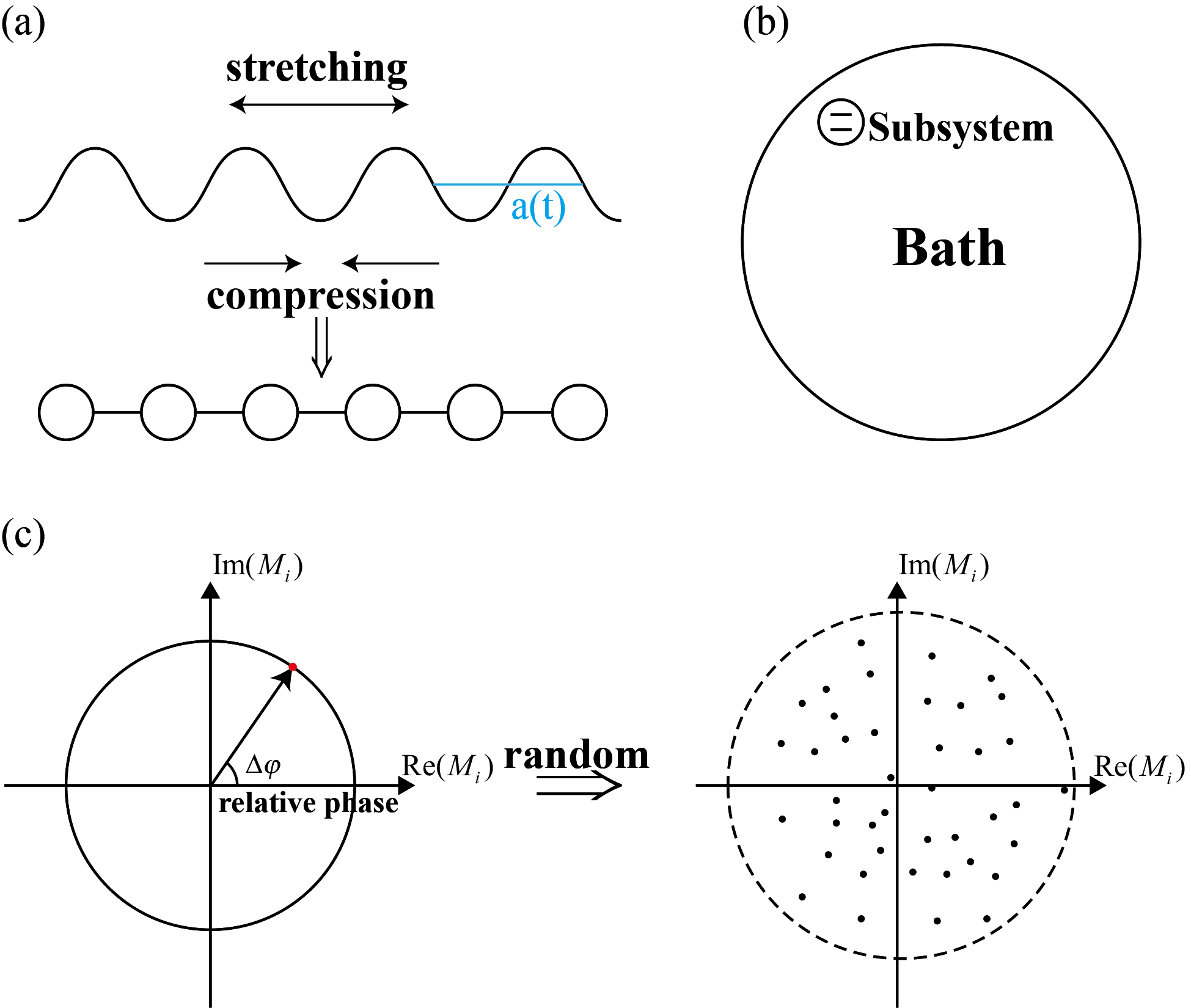}
\label{fig:sub1}
\caption{($a$) Illustration of the accordion-type driving. 
A lattice mode is modulated by a wave whose wavevector varies with time like playing the accordion. 
($b$) Illustration of the structure in the Hilbert space. 
Under the accordion-type driving, there exists a stable subsystem with two states in the Hilbert space and other states can be naturally regarded as the bath around it. 
($c$) Description of randomness in single-particle wave function. $M_{i}=\psi_{i}^{\star}\psi_{L/2+i}$. Every $i \in \{1,2,…,L/2\}$ contributes a point in the graph. If the wave function is a plane wave, all points are the same. In contrast, if the wave function is random, the distribution of points in the diagram will be random.}
 \label{fig-fig1}
\end{figure}

{\it Physical model}. 
Consider the model systems with the following symmetry,
\begin{eqnarray}
H(x,t) &=& H(x+\frac{1}{p f(t)},t), \nonumber \\
H(x, t) &=& H(x, t+T),
\label{eq-general}
\end{eqnarray}
where $f(t)=f(t+T)$ represents the Floquet modulation. 
The new point here is that the translation symmetry becomes time-dependent, yielding a new platform for the interplay between symmetry and modulation, which may lead to a new dynamical structure. 
To this end, we consider the following concrete model, 
\begin{equation}
    H=\sum_{i}V\cos\left[p \sin (\omega t) i\right] c_{i}^{\dagger}c_{i}+hc_{i+1}^{\dagger}c_{i}+hc_{i}^{\dagger}c_{i+1},
    \label{eq-hamil}
\end{equation}
where $h$ is hopping integral set as the energy scale, i.e., 
$h=-1$; $t$ is measured in the unit of $2\pi/h$.
At each instantaneous time $t$, Eq. \ref{eq-hamil} is the same as the Aubry-André (AA) model \cite{articleaah}, yet now the modulation pitch $\alpha=p\sin \omega t$ is time-dependent, which resembles an accordion. 
A schematic illustration of this process 
is presented in Fig. \ref{fig-fig1}($a$). 
It should be noticed that the instantaneous spatial periodicity can be realized in the continuous situation, yet in the discrete lattice model, one may find that when $\alpha$ is incommensurate, the spatial periodicity is lost. 
Nevertheless, this does not affect our major results. 

{\it High frequency limit}. 
We first consider the high frequency limit for insight to the physics in this model. 
By employing the Jacobi-Anger expansion $e^{i A \sin(\omega t)}=\sum_n J_n(A) e^{in\omega t}$, where $J_n(A)$ is the Bessel function of the first kind, we  arrive at
$H=\sum_{n} H_n$ with
\begin{eqnarray}
H_0=\sum_{i}
    \left\{V J_{0}(p i)c_{i}^{\dagger}c_{i}+hc_{i+1}^{\dagger}c_{i}+\mathrm{h.c.} \right\},
    \label{eq-leadingorder}
\end{eqnarray}
and 
\begin{eqnarray}
H_{n (n\neq 0)}=V\sum_i J_{2n}(pi) e^{i2n\omega t} c^\dagger_i c_i.
\label{eq:high}
\end{eqnarray}

In the high frequency limit 
$\omega \gg V,h$, the leading order effective Floquet Hamiltonian is just $H_0$.
By choosing $p=\pi$ and transforming Eq. (\ref{eq-leadingorder}) to momentum space, the leading order 
effective Hamiltonian is,
\begin{equation}
H_{0}(k)=2h\cos k c_{k}^{\dagger}c_{k}+\frac{V}{\pi}\sum_{q}\frac{c_{k+q}^{\dagger}c_{k}+\mathrm{h.c.} }{ \sqrt{1-({q / \pi})^{2}} },
\label{eq-hamilinkspace}
\end{equation}
where $q$ is the scattering momentum and $0\le k\le 2\pi$. 
To arrive at Eq. (\ref{eq-hamilinkspace}),
the integral representation of the Bessel function in the limit of $L\to \infty$ is employed, 
\begin{eqnarray}
    \tilde{J_{n}}(q) = \int_{-\pi}^{\pi} d\theta  e^{in\theta}\delta(q+\pi \sin \theta) 
   = \frac{\exp(-in\sin^{-1}(\frac{q}{\pi}))}
   {\pi\sqrt{1-(\frac{q}{\pi})^{2}}}, 
   \nonumber \\
\label{eq-fourierofbessel}
\end{eqnarray}
where $|q| \le \pi$.  

The singularity of the scattering amplitude 
in Eq. (\ref{eq-fourierofbessel}) naturally distinguishes two regimes: One is the weak scattering regime when $q$ is away from $\pi$, and the other is the strong one as $q \to \pi$.
Due to the band flatness, or, the divergence of
the density of states (DOS) at $k=0$ and $\pi$, if the initial state is a superposition between
$ \{ \ket{k=0}$ and $\ket{k=\pi} \}$, these two states strongly couple spontaneously forming a subsystem.
The other states are treated as a bath.
In this case, the scatterings between the bath and subsystem are weak
as shown in Fig.~\ref{fig-fig1}($b$). 
In contrast, if the initial state is close to 
$|k=\frac{\pi}{2}\rangle$, or, 
$|-\frac{\pi}{2}\rangle$, where the minima of DOS locate, the weak scatterings are dominate leading to a possible fast thermalization. 


\begin{figure}[htbp]
\includegraphics
[width=\columnwidth]{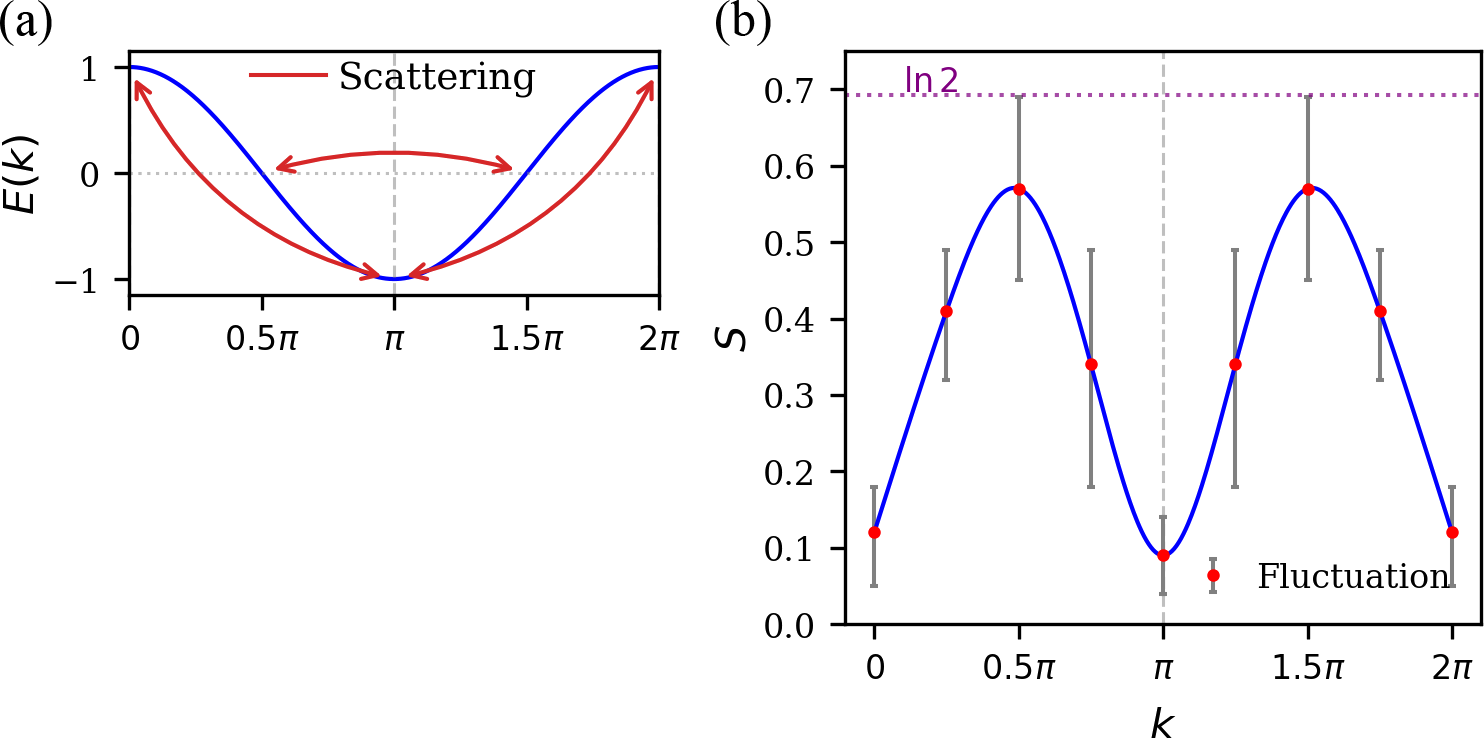}
\caption{($a$) Illustration of the scattering process. The dispersion relation is $E(k)=2h\cos k$. 
The scattering with the wavevector $q=\pm \pi$ is also shown. 
($b$) $S$ with fluctuations at evolution time $t=200$ with the initial state $|k_0\rangle$. The parameters of model are $V=4,h=-1,\omega=10$. The minimal $S$ happens when $k_0=0,\pi$, and the maximal $S$ occurs at $k_0=0.5\pi$ or $1.5\pi$.
The maximal value of $S$ is $\ln 2$.  }
    \label{fig-fig2}
\end{figure}

To quantify the thermalization process, we design a method to characterize the randomness of a single-particle wavefunction.
For a single-particle state $|\psi\rangle=\sum_i \psi_i |i\rangle $
defined in the lattice with $L$ sites
normalized as $\sum_i |\psi_i|^2=1$,
we calculate $M_i=\psi^*_i \psi_{L/2+i}$ and plot their values in the complex plane for $i=1\sim L/2$ as shown in Fig.~\ref{fig-fig1}($c$).
The randomness should reflect in two aspects- the magnitude and phase of $M_i$.
To describe the randomness of $|\psi\rangle$, we define a modified ``entanglement" entropy as
\begin{eqnarray}
S(t)&=&-\mathrm{Tr}[\rho\ln(\rho)],
\label{eq:S}
\end{eqnarray}
with the choice of $\rho$ defined as,
\begin{eqnarray}
\rho&=&\ket{P_A \psi} \bra{P_A \psi}
+\ket{P_A T\psi}\bra{P_A T\psi},
\label{eq-entropy}
\end{eqnarray}
where $A$ is the subsystem with sites from 1 to $L/2$. 
We emphasize that $\rho$ generally does not satisfy the requirement of a density matrix.
$P_A$ is the projection operator defined as $P_A |\psi\rangle =\sum_{i=1}^{L/2} \psi_i |i\rangle $ without normalization.
$T$ is the translation operator which shifts the site index $i\to i+L/2 ~(\mbox{mod}~ L)$.
We define $p_A=\langle\psi|P_A|\psi \rangle $ and $p_B=1-p_A= \langle\psi|T P_A T|\psi \rangle$.

\begin{figure}[htbp]
\includegraphics[width=\columnwidth]{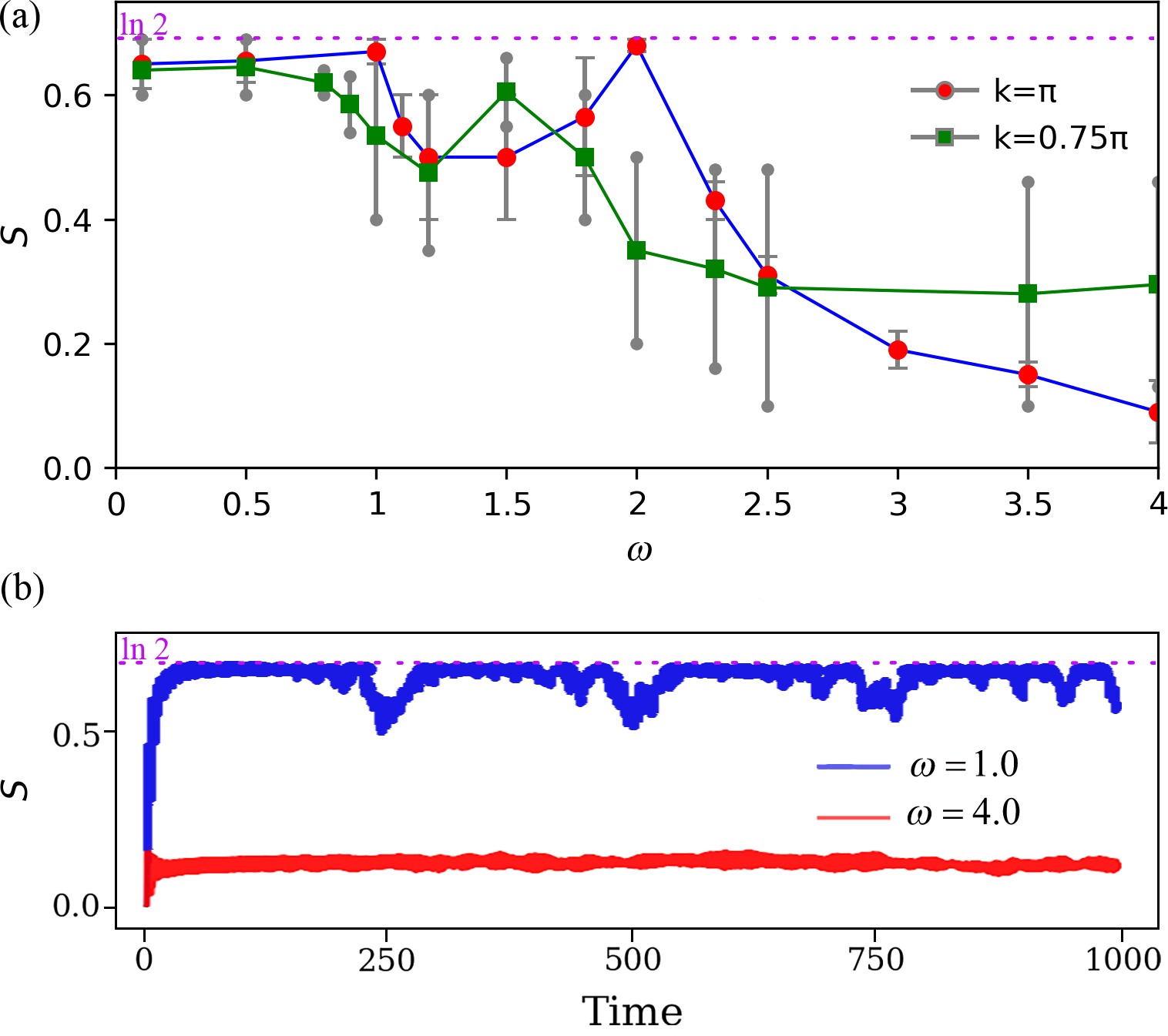}
\caption{($a$) The results of
$S$ defined in Eq. (\ref{eq:S}) after evolving to $t=200$ with 
the initial states 
$|k=\pi\rangle$ and 
$|k=\frac{3}{4}\pi\rangle$ at different values of $\omega$. 
Fluctuations are marked by error bars, and parameter values are $V=4$ and $h=-1$. 
The randomness is enhanced as lowering $\omega$, and tends to reach the maximal value of $\ln 2$.
($b$) The long-time evolutions of $S$  with initial state prepared at $|k=\pi\rangle$ under finite driving frequencies. 
$S$ evolves to equilibrate and keeps stable with small fluctuations in a long time. 
}
\label{fig-fig3}
\end{figure}

Compared with the single-particle entanglement entropy defined in the previous literature \cite{singleparticleentropy}, which is essentially the Shannon entropy.
Our definition based on Eq. (\ref{eq-fourierofbessel}) here contains the phase correlations among sites at large distance. 
If $P_A \ket{\psi}$ is orthogonal to 
$TP_A \ket{\psi}$ and $p_{A}=p_{B}=1/2$, then it means a random distribution of relative phase and probability, $S$ reaches the maximal value $\ln2$. 
In contrast, if $P_A|\psi\rangle = \lambda 
T P_A |\psi\rangle$, or, if $|\psi \rangle$ localizes in the region $A$, then $S = 0$. 
More information on the properties of $S$ 
can be found in the Supplemental Material.
For this reason, we expect this quantity more capable to describe the wavefunction randomness including both aspects of magnitude and phase. 
The typical time-evolution of $S(t)$ is presented in Fig.~\ref{fig-fig3}($b$), which shows that it finally saturates to a certain value with fluctuations. 
The values of $S(t)$ at long-time limit reaching equilibrium in the high-frequency limit are presented in Fig.~\ref{fig-fig2}($b$).
The pair of states $ \{ \ket{k=0},\ket{k=\pi} \}$ indeed form a stable subsystem showing the the minimal value of $S$, while those of $ \{ \ket{k=\frac{\pi}{2}},\ket{k=-\frac{\pi}{2}} \}$ exhibiting the maximal value, {\it i.e.}, 
exhibiting higher level of
randomness. 
 
{\it From fast to slow modulation}. 
Next we study the case that the approximation of $\omega \to \infty$ does not apply
\cite{floquetresonancesclosetotheadiabaticlimit,solvablemodelfordiscretetimecrystalenforcedbynonsymmorphicdynamicalsymmetry,timedependentfloquettheoryandabsenceofanadiabaticlimit}. 
In the picture of photon numbers, a finite frequency results in complicated high-order couplings with multiple photons at different energies, which could destabilize the 
picture of steady states. 
Heating arises from the divergence of the Floquet-Magnus expansion as shown in Ref.  \cite{heatinginintegrabletimeperiodicsystems}. 
In our accordion model, the high-order processes in the Magnus expansion are calculated in detail in the Supplemental Material.
The $1/\omega$-order contribution vanishes, while the $1/\omega^2$ order Floquet-Magnus expansion $H_{\text{eff}}^{(2)}$ results in the divergence in the expansion.
Heating is expected to appear in the low frequency regime.
As shown in Fig. ~\ref{fig-fig3} ($a$), the subsystem evolves to equilibrate exhibiting a low entropy at relatively high frequencies for both cases with initial states $|k=\pi\rangle$
and $|k=\frac{3}{4}\pi\rangle$.
As lowering $\omega$, the system with the initial state $|k=\pi\rangle$ reaches the maximal entropy $\ln 2$ at $\omega\approx 1$ and $2$.
In Fig.~\ref{fig-fig3}($b$), the long-time evolutions of $S$ are shown at $\omega=4$ and $1$, respectively. 
At $\omega=1$, $S$ quickly reaches $\ln 2$ and exhibits certain fluctuations. 


{\it Stable dynamics in the subsystem}. 
Based on the analysis above, we develop a stable subspace consisting of 
$|k=0\rangle$ and $|k=\pi\rangle$ under the accordion-type driving.
We transform Eqs. (\ref{eq-leadingorder}) and  Eq. (\ref{eq:high}) into momentum space and then project to this subspace. 
In order to understand the time-evolution under a finite frequency driving, we focus on
the dynamics within this subspace based on the Floquet theory \cite{perturbativefloquethighfrequency,floquetperturbationtheory},
\begin{equation}
    \ket{\psi(t)}=\sum_{l}a_{l}e^{-i\epsilon_{l}t}\ket{\psi_{l}(t)},
    \label{eq-floquetmode}
\end{equation}
where $\ket{\psi_{l}(t)}$ is a time-dependent eigen-wavefunction with the quasi-energy $\epsilon_{l}$. 
The scattering amplitude at
the wavevector $\pi$ at different orders of $n$ maintains the same amplitude, and exhibits phase difference of $e^{-in\pi/2}$ according to
Eq. (\ref{eq-fourierofbessel}).
A phenomenological parameter $\tilde{V}$ is introduced here to describe this singular scattering between $\ket{k=0}$ and $\ket{k=\pi}$. 
Hence, the effective Hamiltonian shows a 
kicked structure as
\begin{equation}
\begin{split}
    H_{\text{sub}}(t)&=H_{s}+\sum_{n}\tilde{V}_{s}e^{-in\pi}e^{-i2n\omega t}\\
    &= H_{s}+2\pi\sum_{n}\tilde{V}_{s}\delta\left(t-(n-\frac{1}{2}) \frac{\pi}{\omega}\right),
\end{split}
\label{eq-kickhamil}
\end{equation}
where $H_{s}$ and $\tilde{V}_{s}$ takes the form 
\begin{equation}
H_{s}=
\begin{bmatrix}
    2h  &  0 \\
    0 &-2h
\end{bmatrix}
,\tilde{V}_{s}=\begin{bmatrix}
    0 & \tilde{V} \\
    \tilde{V} & 0
\end{bmatrix}.
\label{eq-2matrix}
\end{equation}

This kicked-type Hamiltonian of Eq. (\ref{eq-kickhamil}) can be investigated by the 1st order perturbative theory, yielding 
\begin{equation}
\begin{split}
    \ket{\psi_{1}(t)}&\approx\ket{k=0}-\tilde{V}
    \sum_{n} (-1)^n \frac{ e^{-i2n \omega t}}{2n\omega-4h}\ket{k=\pi},
\end{split}
\label{eq-perturbativewave}
\end{equation}
which explains the behavior of $S$. When $\omega/h=2$, the perturbative series diverge, causing the peak of $S$ as shown in Fig. \ref{fig-fig3} ($a$).
Furthermore, when $\omega/h=4/(2m+1)$ with $m$ an integer, termed as the square wave condition below, the summation in Eq. (\ref{eq-perturbativewave}) can be exactly performed. 
Under the condition that $\frac{\pi \tilde{V}}{2\omega}\ll 1$, the Floquet eigen wavefunctions are expressed as,
\begin{equation}
\begin{split}
    \ket{\psi_{1}(t)}&\approx\ket{k=0}+\frac{2\tilde{V}}{\omega}g(t)e^{-i(2m+1)\omega t}
    \ket{k=\pi},\\
    \ket{\psi_{2}(t)} &\approx\ket{k=\pi}-\frac{2\tilde{V}}{\omega}g(t)e^{i(2m+1)\omega t}
    \ket{k=0},
\end{split}
\label{eq-eigenwavefunction}
\end{equation}
where 
\begin{equation}
g(t)=\sum_{n=0}^{\infty}
\frac{\cos\left((2n+1\right)\omega t)}{2n+1}
 =\frac{\pi}{4}\text{sgn}\left(\cos \omega t \right),
\label{eq-series}
\end{equation}
representing a square wave. 
For example, if $\omega/h=-4/3$, the square wave condition causes the suppression of $S$ around $\omega=\frac{4}{3}$.

\begin{figure}[htbp]
\includegraphics[width=\columnwidth]{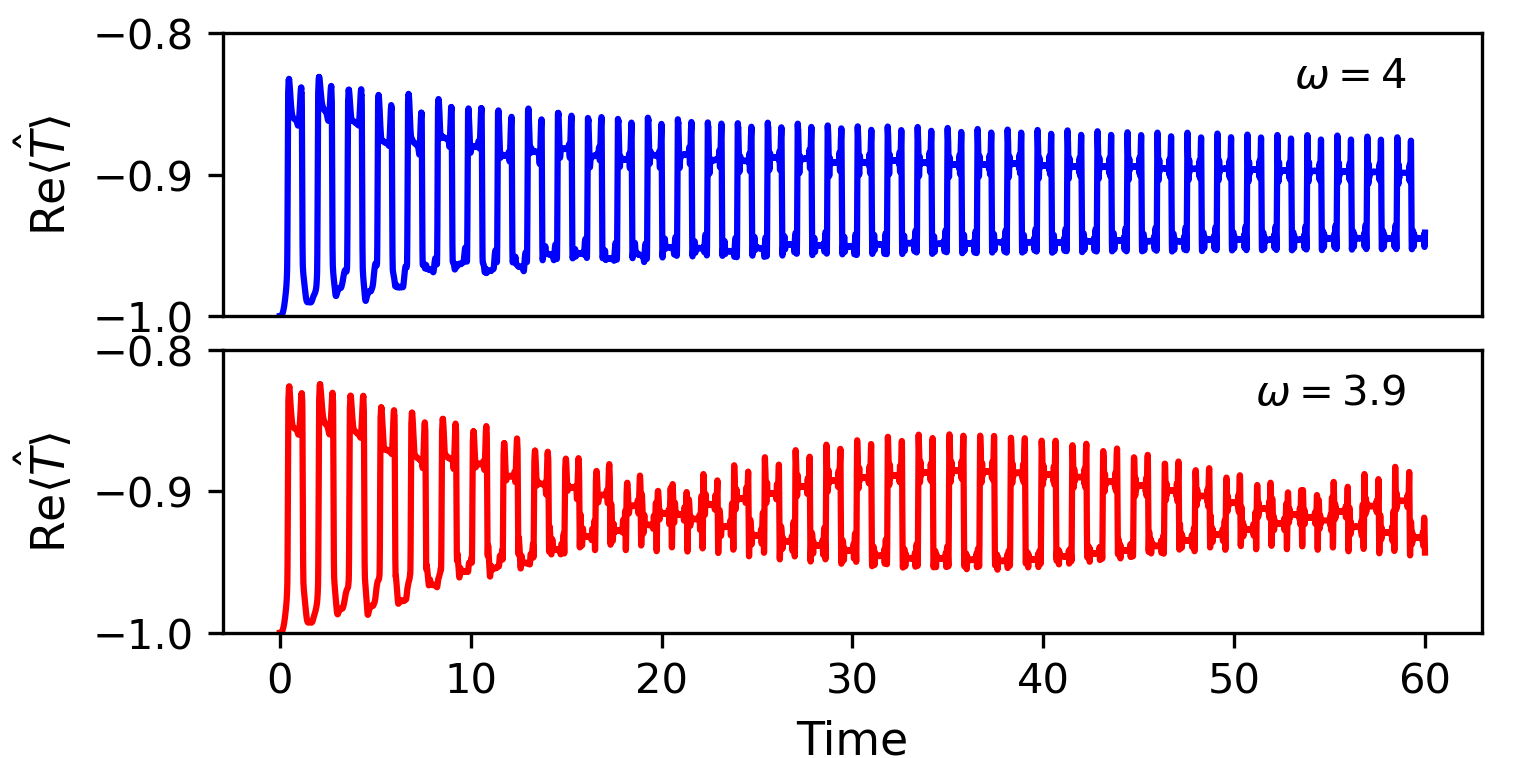}
\caption{The evolution of the expectation value of translational operator $\hat T$ with the initial wavefunction $\ket{k=\pi}$. 
After a loading process, there is a stable square wave behavior in a period $2\pi/\omega$. 
A little deviation of $\omega=4$ will cause a standing wave mode with respect to time. 
The peaks around filps are the Gibbs phenomenon \cite{Gibbsphenomenon}, which can be suppressed when increasing the size of lattice as shown in Supplemental Material. 
}
\label{fig-fig4}
\end{figure}

Next we present the intuitive results of the square wave condition.
$\hat{T}$ is defined as the operator translating one lattice constant, satisfying
$\hat{T}\ket{k=0(\pi)}=\pm \ket{k=0(\pi)}$.
Choose the initial state 
$|\psi(t=0)\rangle = a_1 |\psi_1(t=0)\rangle +
a_2 |\psi_2(t=0)\rangle$.
Under the condition that $|a_{1}|\ll |a_{2}|$, the expectation value of $\hat{T}$ is calculated 
\begin{equation}
\langle\psi(t) |\hat{T} | \psi(t) \rangle 
\approx -a_{2}^{2}-(a_{1}a_{2}^{\star}+a_{1}^{\star}a_{2})\frac{\pi\tilde{V}}{\omega} f(t).
    \label{eq-stationaryeigen}
\end{equation}
By choosing 
$|\psi(t=0)\rangle=|k=\pi \rangle$
, then $a_1/a_2\approx 2\tilde{V}/\omega\ll 1$, we should arrive at
a square-wave like evolution, which  is numerically  at $\omega=-4h$ as shown in Fig.~\ref{fig-fig4}.
This shape can be regarded as a modification of the two-level oscillation originating from the coupling with photons of energy $n\omega$.
More details about the derivation of these results and discussions of the perturbative series can be found in the Supplemental Material.

{\it Conclusion and discussion}.
We have studied the quantum dynamics of a Floquet system under
the accordion-type driving.
It is found that the Hilbert space is divided into two subspaces, one of which consists of two states, playing the role of two-level system and the rest part behaves like a bath. 
We demonstrate that the effective two-level system exhibit stable and controllable dynamical characters for a very long time. 
We hope that this kind of spontaneous formation of a stable two-level subsystem is useful for further exploring many-body physics like quantum Scar states 
due to their similarities \cite{quantummanybodyscars}. 
At the same time, we propose a general method to investigate the kicked structure with perturbative time-dependent eigen wave function, which can be used to investigate a wide range of kicked potentials. 
This work enriches our understanding of kicked evolution and the quantum thermalization process by modulation.   

Since this design of singularity in the scattering does not depend on the precise form of periodic modulation $f(t)$, we expect that the related experiments in the ultracold atoms are achievable and one possible construction is introduced in the Supplemental Material, which can be regarded as rotating a two-dimensional standing wave field on a one-dimensional tight-binding chain.  

{\it Acknowledgments}.
C. W. is supported by the National Natural Science Foundation of China (NSFC) under the Grant No. 12234016, and also supported by the NSFC under the Grant Nos. 12174317, respectively.  This work has been supported by the New Cornerstone Science Foundation.
\bibliography{references} 

\begin{thebibliography}{1}

\bibitem{singleparticleentropy}
Faris Abualnaja, Wenkun He, Aleksey Andreev, Mervyn Jones, and Zahid Durrani.
\newblock Single particle entropy stability and the temperature-entropy diagram in quantum dot transistors.
\newblock {\em Phys. Rev. Res.}, 5:033025, Jul 2023.

\bibitem{Bukov_review}
Marin Bukov, Luca D’Alessio, and Anatoli Polkovnikov.
\newblock Universal high-frequency behavior of periodically driven systems: from dynamical stabilization to floquet engineering.
\newblock {\em Advances in Physics}, 64(2):139–226, March 2015.

\bibitem{Kuwahara_2016floquetmagnustheory}
Tomotaka Kuwahara, Takashi Mori, and Keiji Saito.
\newblock Floquet–magnus theory and generic transient dynamics in periodically driven many-body quantum systems.
\newblock {\em Annals of Physics}, 367:96–124, April 2016.

\bibitem{spacetimecrystal}
Shenglong Xu and Congjun Wu.
\newblock Space-time crystal and space-time group.
\newblock {\em Phys. Rev. Lett.}, 120:096401, Feb 2018.

\bibitem{perturbativefloquethighfrequency}
André Eckardt and Egidijus Anisimovas.
\newblock High-frequency approximation for periodically driven quantum systems from a floquet-space perspective.
\newblock {\em New Journal of Physics}, 17(9):093039, September 2015.

\bibitem{floquetperturbationtheory}
M~Rodriguez-Vega, M~Lentz, and B~Seradjeh.
\newblock Floquet perturbation theory: formalism and application to low-frequency limit.
\newblock {\em New Journal of Physics}, 20(9):093022, sep 2018.

\end{thebibliography}


\begin{thebibliography}{36}%
\makeatletter
\providecommand \@ifxundefined [1]{%
 \@ifx{#1\undefined}
}%
\providecommand \@ifnum [1]{%
 \ifnum #1\expandafter \@firstoftwo
 \else \expandafter \@secondoftwo
 \fi
}%
\providecommand \@ifx [1]{%
 \ifx #1\expandafter \@firstoftwo
 \else \expandafter \@secondoftwo
 \fi
}%
\providecommand \natexlab [1]{#1}%
\providecommand \enquote  [1]{``#1''}%
\providecommand \bibnamefont  [1]{#1}%
\providecommand \bibfnamefont [1]{#1}%
\providecommand \citenamefont [1]{#1}%
\providecommand \href@noop [0]{\@secondoftwo}%
\providecommand \href [0]{\begingroup \@sanitize@url \@href}%
\providecommand \@href[1]{\@@startlink{#1}\@@href}%
\providecommand \@@href[1]{\endgroup#1\@@endlink}%
\providecommand \@sanitize@url [0]{\catcode `\\12\catcode `\$12\catcode `\&12\catcode `\#12\catcode `\^12\catcode `\_12\catcode `\%12\relax}%
\providecommand \@@startlink[1]{}%
\providecommand \@@endlink[0]{}%
\providecommand \url  [0]{\begingroup\@sanitize@url \@url }%
\providecommand \@url [1]{\endgroup\@href {#1}{\urlprefix }}%
\providecommand \urlprefix  [0]{URL }%
\providecommand \Eprint [0]{\href }%
\providecommand \doibase [0]{https://doi.org/}%
\providecommand \selectlanguage [0]{\@gobble}%
\providecommand \bibinfo  [0]{\@secondoftwo}%
\providecommand \bibfield  [0]{\@secondoftwo}%
\providecommand \translation [1]{[#1]}%
\providecommand \BibitemOpen [0]{}%
\providecommand \bibitemStop [0]{}%
\providecommand \bibitemNoStop [0]{.\EOS\space}%
\providecommand \EOS [0]{\spacefactor3000\relax}%
\providecommand \BibitemShut  [1]{\csname bibitem#1\endcsname}%
\let\auto@bib@innerbib\@empty
\bibitem [{\citenamefont {Merboldt}\ \emph {et~al.}(2025)\citenamefont {Merboldt}, \citenamefont {Schüler}, \citenamefont {Schmitt}, \citenamefont {Bange}, \citenamefont {Bennecke}, \citenamefont {Gadge}, \citenamefont {Pierz}, \citenamefont {Schumacher}, \citenamefont {Momeni}, \citenamefont {Steil}, \citenamefont {Manmana}, \citenamefont {Sentef}, \citenamefont {Reutzel},\ and\ \citenamefont {Mathias}}]{observationoffloquetstatesingraphene}%
  \BibitemOpen
  \bibfield  {author} {\bibinfo {author} {\bibfnamefont {M.}~\bibnamefont {Merboldt}}, \bibinfo {author} {\bibfnamefont {M.}~\bibnamefont {Schüler}}, \bibinfo {author} {\bibfnamefont {D.}~\bibnamefont {Schmitt}}, \bibinfo {author} {\bibfnamefont {J.~P.}\ \bibnamefont {Bange}}, \bibinfo {author} {\bibfnamefont {W.}~\bibnamefont {Bennecke}}, \bibinfo {author} {\bibfnamefont {K.}~\bibnamefont {Gadge}}, \bibinfo {author} {\bibfnamefont {K.}~\bibnamefont {Pierz}}, \bibinfo {author} {\bibfnamefont {H.~W.}\ \bibnamefont {Schumacher}}, \bibinfo {author} {\bibfnamefont {D.}~\bibnamefont {Momeni}}, \bibinfo {author} {\bibfnamefont {D.}~\bibnamefont {Steil}}, \bibinfo {author} {\bibfnamefont {S.~R.}\ \bibnamefont {Manmana}}, \bibinfo {author} {\bibfnamefont {M.~A.}\ \bibnamefont {Sentef}}, \bibinfo {author} {\bibfnamefont {M.}~\bibnamefont {Reutzel}},\ and\ \bibinfo {author} {\bibfnamefont {S.}~\bibnamefont {Mathias}},\ }\bibfield  {title} {\bibinfo {title} {Observation of floquet states in graphene},\ }\href
  {https://doi.org/10.1038/s41567-025-02889-7} {\bibfield  {journal} {\bibinfo  {journal} {Nature Physics}\ } (\bibinfo {year} {2025})}\BibitemShut {NoStop}%
\bibitem [{\citenamefont {Oka}\ and\ \citenamefont {Kitamura}(2019)}]{floquetengineeringofquantummaterials}%
  \BibitemOpen
  \bibfield  {author} {\bibinfo {author} {\bibfnamefont {T.}~\bibnamefont {Oka}}\ and\ \bibinfo {author} {\bibfnamefont {S.}~\bibnamefont {Kitamura}},\ }\bibfield  {title} {\bibinfo {title} {Floquet engineering of quantum materials},\ }\href {https://doi.org/https://doi.org/10.1146/annurev-conmatphys-031218-013423} {\bibfield  {journal} {\bibinfo  {journal} {Annual Review of Condensed Matter Physics}\ }\textbf {\bibinfo {volume} {10}},\ \bibinfo {pages} {387} (\bibinfo {year} {2019})}\BibitemShut {NoStop}%
\bibitem [{\citenamefont {Aeschlimann}\ \emph {et~al.}(2021)\citenamefont {Aeschlimann}, \citenamefont {Sato}, \citenamefont {Krause}, \citenamefont {Chávez-Cervantes}, \citenamefont {De~Giovannini}, \citenamefont {Hübener}, \citenamefont {Forti}, \citenamefont {Coletti}, \citenamefont {Hanff}, \citenamefont {Rossnagel}, \citenamefont {Rubio},\ and\ \citenamefont {Gierz}}]{condensedmatterexsurvivalofFBstatesinthepresenceofscattering}%
  \BibitemOpen
  \bibfield  {author} {\bibinfo {author} {\bibfnamefont {S.}~\bibnamefont {Aeschlimann}}, \bibinfo {author} {\bibfnamefont {S.~A.}\ \bibnamefont {Sato}}, \bibinfo {author} {\bibfnamefont {R.}~\bibnamefont {Krause}}, \bibinfo {author} {\bibfnamefont {M.}~\bibnamefont {Chávez-Cervantes}}, \bibinfo {author} {\bibfnamefont {U.}~\bibnamefont {De~Giovannini}}, \bibinfo {author} {\bibfnamefont {H.}~\bibnamefont {Hübener}}, \bibinfo {author} {\bibfnamefont {S.}~\bibnamefont {Forti}}, \bibinfo {author} {\bibfnamefont {C.}~\bibnamefont {Coletti}}, \bibinfo {author} {\bibfnamefont {K.}~\bibnamefont {Hanff}}, \bibinfo {author} {\bibfnamefont {K.}~\bibnamefont {Rossnagel}}, \bibinfo {author} {\bibfnamefont {A.}~\bibnamefont {Rubio}},\ and\ \bibinfo {author} {\bibfnamefont {I.}~\bibnamefont {Gierz}},\ }\bibfield  {title} {\bibinfo {title} {Survival of floquet–bloch states in the presence of scattering},\ }\href {https://doi.org/10.1021/acs.nanolett.1c00801} {\bibfield  {journal} {\bibinfo  {journal} {Nano Letters}\
  }\textbf {\bibinfo {volume} {21}},\ \bibinfo {pages} {5028–5035} (\bibinfo {year} {2021})}\BibitemShut {NoStop}%
\bibitem [{\citenamefont {Singh}\ \emph {et~al.}(2019)\citenamefont {Singh}, \citenamefont {Fujiwara}, \citenamefont {Geiger}, \citenamefont {Simmons}, \citenamefont {Lipatov}, \citenamefont {Cao}, \citenamefont {Dotti}, \citenamefont {Rajagopal}, \citenamefont {Senaratne}, \citenamefont {Shimasaki}, \citenamefont {Heyl}, \citenamefont {Eckardt},\ and\ \citenamefont {Weld}}]{experimentalprethermalfloquet}%
  \BibitemOpen
  \bibfield  {author} {\bibinfo {author} {\bibfnamefont {K.}~\bibnamefont {Singh}}, \bibinfo {author} {\bibfnamefont {C.~J.}\ \bibnamefont {Fujiwara}}, \bibinfo {author} {\bibfnamefont {Z.~A.}\ \bibnamefont {Geiger}}, \bibinfo {author} {\bibfnamefont {E.~Q.}\ \bibnamefont {Simmons}}, \bibinfo {author} {\bibfnamefont {M.}~\bibnamefont {Lipatov}}, \bibinfo {author} {\bibfnamefont {A.}~\bibnamefont {Cao}}, \bibinfo {author} {\bibfnamefont {P.}~\bibnamefont {Dotti}}, \bibinfo {author} {\bibfnamefont {S.~V.}\ \bibnamefont {Rajagopal}}, \bibinfo {author} {\bibfnamefont {R.}~\bibnamefont {Senaratne}}, \bibinfo {author} {\bibfnamefont {T.}~\bibnamefont {Shimasaki}}, \bibinfo {author} {\bibfnamefont {M.}~\bibnamefont {Heyl}}, \bibinfo {author} {\bibfnamefont {A.}~\bibnamefont {Eckardt}},\ and\ \bibinfo {author} {\bibfnamefont {D.~M.}\ \bibnamefont {Weld}},\ }\bibfield  {title} {\bibinfo {title} {Quantifying and controlling prethermal nonergodicity in interacting floquet matter},\ }\href
  {https://doi.org/10.1103/PhysRevX.9.041021} {\bibfield  {journal} {\bibinfo  {journal} {Phys. Rev. X}\ }\textbf {\bibinfo {volume} {9}},\ \bibinfo {pages} {041021} (\bibinfo {year} {2019})}\BibitemShut {NoStop}%
\bibitem [{\citenamefont {Dotti}\ \emph {et~al.}(2024)\citenamefont {Dotti}, \citenamefont {Bai}, \citenamefont {Shimasaki}, \citenamefont {Dardia},\ and\ \citenamefont {Weld}}]{dotti2024measuringlocalizationphasediagram}%
  \BibitemOpen
  \bibfield  {author} {\bibinfo {author} {\bibfnamefont {P.}~\bibnamefont {Dotti}}, \bibinfo {author} {\bibfnamefont {Y.}~\bibnamefont {Bai}}, \bibinfo {author} {\bibfnamefont {T.}~\bibnamefont {Shimasaki}}, \bibinfo {author} {\bibfnamefont {A.~R.}\ \bibnamefont {Dardia}},\ and\ \bibinfo {author} {\bibfnamefont {D.~M.}\ \bibnamefont {Weld}},\ }\href {https://arxiv.org/abs/2406.00214} {\bibinfo {title} {Measuring a localization phase diagram controlled by the interplay of disorder and driving}} (\bibinfo {year} {2024}),\ \Eprint {https://arxiv.org/abs/2406.00214} {arXiv:2406.00214 [physics.atom-ph]} \BibitemShut {NoStop}%
\bibitem [{\citenamefont {Bukov}\ \emph {et~al.}(2015)\citenamefont {Bukov}, \citenamefont {D’Alessio},\ and\ \citenamefont {Polkovnikov}}]{Bukov_review}%
  \BibitemOpen
  \bibfield  {author} {\bibinfo {author} {\bibfnamefont {M.}~\bibnamefont {Bukov}}, \bibinfo {author} {\bibfnamefont {L.}~\bibnamefont {D’Alessio}},\ and\ \bibinfo {author} {\bibfnamefont {A.}~\bibnamefont {Polkovnikov}},\ }\bibfield  {title} {\bibinfo {title} {Universal high-frequency behavior of periodically driven systems: from dynamical stabilization to floquet engineering},\ }\href {https://doi.org/10.1080/00018732.2015.1055918} {\bibfield  {journal} {\bibinfo  {journal} {Advances in Physics}\ }\textbf {\bibinfo {volume} {64}},\ \bibinfo {pages} {139–226} (\bibinfo {year} {2015})}\BibitemShut {NoStop}%
\bibitem [{\citenamefont {Fu}\ \emph {et~al.}(2024)\citenamefont {Fu}, \citenamefont {Moessner}, \citenamefont {Zhao},\ and\ \citenamefont {Bukov}}]{floquethierarchicalsymmetry}%
  \BibitemOpen
  \bibfield  {author} {\bibinfo {author} {\bibfnamefont {Z.}~\bibnamefont {Fu}}, \bibinfo {author} {\bibfnamefont {R.}~\bibnamefont {Moessner}}, \bibinfo {author} {\bibfnamefont {H.}~\bibnamefont {Zhao}},\ and\ \bibinfo {author} {\bibfnamefont {M.}~\bibnamefont {Bukov}},\ }\bibfield  {title} {\bibinfo {title} {Engineering hierarchical symmetries},\ }\bibfield  {journal} {\bibinfo  {journal} {Physical Review X}\ }\textbf {\bibinfo {volume} {14}},\ \href {https://doi.org/10.1103/physrevx.14.041070} {10.1103/physrevx.14.041070} (\bibinfo {year} {2024})\BibitemShut {NoStop}%
\bibitem [{\citenamefont {Eckardt}(2017)}]{atomicquantumgasesinperiodicallydrivenopticallattices31}%
  \BibitemOpen
  \bibfield  {author} {\bibinfo {author} {\bibfnamefont {A.}~\bibnamefont {Eckardt}},\ }\bibfield  {title} {\bibinfo {title} {Colloquium: Atomic quantum gases in periodically driven optical lattices},\ }\href {https://doi.org/10.1103/RevModPhys.89.011004} {\bibfield  {journal} {\bibinfo  {journal} {Rev. Mod. Phys.}\ }\textbf {\bibinfo {volume} {89}},\ \bibinfo {pages} {011004} (\bibinfo {year} {2017})}\BibitemShut {NoStop}%
\bibitem [{\citenamefont {Jotzu}\ \emph {et~al.}(2014)\citenamefont {Jotzu}, \citenamefont {Messer}, \citenamefont {Desbuquois}, \citenamefont {Lebrat}, \citenamefont {Uehlinger}, \citenamefont {Greif},\ and\ \citenamefont {Esslinger}}]{experimentalrealizationofhaldanemodel}%
  \BibitemOpen
  \bibfield  {author} {\bibinfo {author} {\bibfnamefont {G.}~\bibnamefont {Jotzu}}, \bibinfo {author} {\bibfnamefont {M.}~\bibnamefont {Messer}}, \bibinfo {author} {\bibfnamefont {R.}~\bibnamefont {Desbuquois}}, \bibinfo {author} {\bibfnamefont {M.}~\bibnamefont {Lebrat}}, \bibinfo {author} {\bibfnamefont {T.}~\bibnamefont {Uehlinger}}, \bibinfo {author} {\bibfnamefont {D.}~\bibnamefont {Greif}},\ and\ \bibinfo {author} {\bibfnamefont {T.}~\bibnamefont {Esslinger}},\ }\bibfield  {title} {\bibinfo {title} {Experimental realization of the topological haldane model with ultracold fermions},\ }\href {https://doi.org/10.1038/nature13915} {\bibfield  {journal} {\bibinfo  {journal} {Nature}\ }\textbf {\bibinfo {volume} {515}},\ \bibinfo {pages} {237–240} (\bibinfo {year} {2014})}\BibitemShut {NoStop}%
\bibitem [{\citenamefont {Miyake}\ \emph {et~al.}(2013)\citenamefont {Miyake}, \citenamefont {Siviloglou}, \citenamefont {Kennedy}, \citenamefont {Burton},\ and\ \citenamefont {Ketterle}}]{realizingtheharperhamiltonianinopticallattices}%
  \BibitemOpen
  \bibfield  {author} {\bibinfo {author} {\bibfnamefont {H.}~\bibnamefont {Miyake}}, \bibinfo {author} {\bibfnamefont {G.~A.}\ \bibnamefont {Siviloglou}}, \bibinfo {author} {\bibfnamefont {C.~J.}\ \bibnamefont {Kennedy}}, \bibinfo {author} {\bibfnamefont {W.~C.}\ \bibnamefont {Burton}},\ and\ \bibinfo {author} {\bibfnamefont {W.}~\bibnamefont {Ketterle}},\ }\bibfield  {title} {\bibinfo {title} {Realizing the harper hamiltonian with laser-assisted tunneling in optical lattices},\ }\href {https://doi.org/10.1103/PhysRevLett.111.185302} {\bibfield  {journal} {\bibinfo  {journal} {Phys. Rev. Lett.}\ }\textbf {\bibinfo {volume} {111}},\ \bibinfo {pages} {185302} (\bibinfo {year} {2013})}\BibitemShut {NoStop}%
\bibitem [{\citenamefont {Atala}\ \emph {et~al.}(2014)\citenamefont {Atala}, \citenamefont {Aidelsburger}, \citenamefont {Lohse}, \citenamefont {Barreiro}, \citenamefont {Paredes},\ and\ \citenamefont {Bloch}}]{chiralcurrents}%
  \BibitemOpen
  \bibfield  {author} {\bibinfo {author} {\bibfnamefont {M.}~\bibnamefont {Atala}}, \bibinfo {author} {\bibfnamefont {M.}~\bibnamefont {Aidelsburger}}, \bibinfo {author} {\bibfnamefont {M.}~\bibnamefont {Lohse}}, \bibinfo {author} {\bibfnamefont {J.~T.}\ \bibnamefont {Barreiro}}, \bibinfo {author} {\bibfnamefont {B.}~\bibnamefont {Paredes}},\ and\ \bibinfo {author} {\bibfnamefont {I.}~\bibnamefont {Bloch}},\ }\bibfield  {title} {\bibinfo {title} {Observation of chiral currents with ultracold atoms in bosonic ladders},\ }\href {https://doi.org/10.1038/nphys2998} {\bibfield  {journal} {\bibinfo  {journal} {Nature Physics}\ }\textbf {\bibinfo {volume} {10}},\ \bibinfo {pages} {588} (\bibinfo {year} {2014})}\BibitemShut {NoStop}%
\bibitem [{\citenamefont {Rechtsman}\ \emph {et~al.}(2013)\citenamefont {Rechtsman}, \citenamefont {Zeuner}, \citenamefont {Plotnik}, \citenamefont {Lumer}, \citenamefont {Podolsky}, \citenamefont {Dreisow}, \citenamefont {Nolte}, \citenamefont {Segev},\ and\ \citenamefont {Szameit}}]{photonicfloquettopologicalinsulators}%
  \BibitemOpen
  \bibfield  {author} {\bibinfo {author} {\bibfnamefont {M.~C.}\ \bibnamefont {Rechtsman}}, \bibinfo {author} {\bibfnamefont {J.~M.}\ \bibnamefont {Zeuner}}, \bibinfo {author} {\bibfnamefont {Y.}~\bibnamefont {Plotnik}}, \bibinfo {author} {\bibfnamefont {Y.}~\bibnamefont {Lumer}}, \bibinfo {author} {\bibfnamefont {D.}~\bibnamefont {Podolsky}}, \bibinfo {author} {\bibfnamefont {F.}~\bibnamefont {Dreisow}}, \bibinfo {author} {\bibfnamefont {S.}~\bibnamefont {Nolte}}, \bibinfo {author} {\bibfnamefont {M.}~\bibnamefont {Segev}},\ and\ \bibinfo {author} {\bibfnamefont {A.}~\bibnamefont {Szameit}},\ }\bibfield  {title} {\bibinfo {title} {Photonic floquet topological insulators},\ }\href {https://doi.org/10.1038/nature12066} {\bibfield  {journal} {\bibinfo  {journal} {Nature}\ }\textbf {\bibinfo {volume} {496}},\ \bibinfo {pages} {196} (\bibinfo {year} {2013})}\BibitemShut {NoStop}%
\bibitem [{\citenamefont {Feng}\ \emph {et~al.}(2014)\citenamefont {Feng}, \citenamefont {Wang}, \citenamefont {Wei}, \citenamefont {Chen}, \citenamefont {Simon}, \citenamefont {Arvapally}, \citenamefont {Martin}, \citenamefont {Bosch}, \citenamefont {Liu}, \citenamefont {Fordham}, \citenamefont {Yuan}, \citenamefont {Omary}, \citenamefont {Haranczyk}, \citenamefont {Smit},\ and\ \citenamefont {Zhou}}]{chiralspinliquidphase}%
  \BibitemOpen
  \bibfield  {author} {\bibinfo {author} {\bibfnamefont {D.}~\bibnamefont {Feng}}, \bibinfo {author} {\bibfnamefont {K.}~\bibnamefont {Wang}}, \bibinfo {author} {\bibfnamefont {Z.}~\bibnamefont {Wei}}, \bibinfo {author} {\bibfnamefont {Y.-P.}\ \bibnamefont {Chen}}, \bibinfo {author} {\bibfnamefont {C.~M.}\ \bibnamefont {Simon}}, \bibinfo {author} {\bibfnamefont {R.~K.}\ \bibnamefont {Arvapally}}, \bibinfo {author} {\bibfnamefont {R.~L.}\ \bibnamefont {Martin}}, \bibinfo {author} {\bibfnamefont {M.}~\bibnamefont {Bosch}}, \bibinfo {author} {\bibfnamefont {T.-F.}\ \bibnamefont {Liu}}, \bibinfo {author} {\bibfnamefont {S.}~\bibnamefont {Fordham}}, \bibinfo {author} {\bibfnamefont {D.}~\bibnamefont {Yuan}}, \bibinfo {author} {\bibfnamefont {M.~A.}\ \bibnamefont {Omary}}, \bibinfo {author} {\bibfnamefont {M.}~\bibnamefont {Haranczyk}}, \bibinfo {author} {\bibfnamefont {B.}~\bibnamefont {Smit}},\ and\ \bibinfo {author} {\bibfnamefont {H.-C.}\ \bibnamefont {Zhou}},\ }\bibfield  {title} {\bibinfo {title} {Kinetically
  tuned dimensional augmentation as a versatile synthetic route towards robust metal--organic frameworks},\ }\href {https://doi.org/10.1038/ncomms6723} {\bibfield  {journal} {\bibinfo  {journal} {Nature Communications}\ }\textbf {\bibinfo {volume} {5}},\ \bibinfo {pages} {5723} (\bibinfo {year} {2014})}\BibitemShut {NoStop}%
\bibitem [{\citenamefont {Struck}\ \emph {et~al.}(2012)\citenamefont {Struck}, \citenamefont {\"Olschl\"ager}, \citenamefont {Weinberg}, \citenamefont {Hauke}, \citenamefont {Simonet}, \citenamefont {Eckardt}, \citenamefont {Lewenstein}, \citenamefont {Sengstock},\ and\ \citenamefont {Windpassinger}}]{tunablegaugepotential}%
  \BibitemOpen
  \bibfield  {author} {\bibinfo {author} {\bibfnamefont {J.}~\bibnamefont {Struck}}, \bibinfo {author} {\bibfnamefont {C.}~\bibnamefont {\"Olschl\"ager}}, \bibinfo {author} {\bibfnamefont {M.}~\bibnamefont {Weinberg}}, \bibinfo {author} {\bibfnamefont {P.}~\bibnamefont {Hauke}}, \bibinfo {author} {\bibfnamefont {J.}~\bibnamefont {Simonet}}, \bibinfo {author} {\bibfnamefont {A.}~\bibnamefont {Eckardt}}, \bibinfo {author} {\bibfnamefont {M.}~\bibnamefont {Lewenstein}}, \bibinfo {author} {\bibfnamefont {K.}~\bibnamefont {Sengstock}},\ and\ \bibinfo {author} {\bibfnamefont {P.}~\bibnamefont {Windpassinger}},\ }\bibfield  {title} {\bibinfo {title} {Tunable gauge potential for neutral and spinless particles in driven optical lattices},\ }\href {https://doi.org/10.1103/PhysRevLett.108.225304} {\bibfield  {journal} {\bibinfo  {journal} {Phys. Rev. Lett.}\ }\textbf {\bibinfo {volume} {108}},\ \bibinfo {pages} {225304} (\bibinfo {year} {2012})}\BibitemShut {NoStop}%
\bibitem [{\citenamefont {Kuwahara}\ \emph {et~al.}(2016)\citenamefont {Kuwahara}, \citenamefont {Mori},\ and\ \citenamefont {Saito}}]{Kuwahara_2016floquetmagnustheory}%
  \BibitemOpen
  \bibfield  {author} {\bibinfo {author} {\bibfnamefont {T.}~\bibnamefont {Kuwahara}}, \bibinfo {author} {\bibfnamefont {T.}~\bibnamefont {Mori}},\ and\ \bibinfo {author} {\bibfnamefont {K.}~\bibnamefont {Saito}},\ }\bibfield  {title} {\bibinfo {title} {Floquet–magnus theory and generic transient dynamics in periodically driven many-body quantum systems},\ }\href {https://doi.org/10.1016/j.aop.2016.01.012} {\bibfield  {journal} {\bibinfo  {journal} {Annals of Physics}\ }\textbf {\bibinfo {volume} {367}},\ \bibinfo {pages} {96–124} (\bibinfo {year} {2016})}\BibitemShut {NoStop}%
\bibitem [{\citenamefont {D'Alessio}\ and\ \citenamefont {Rigol}(2014)}]{longtimebehaviorofisolatedperiodicallydriveninteractinglatticesystems}%
  \BibitemOpen
  \bibfield  {author} {\bibinfo {author} {\bibfnamefont {L.}~\bibnamefont {D'Alessio}}\ and\ \bibinfo {author} {\bibfnamefont {M.}~\bibnamefont {Rigol}},\ }\bibfield  {title} {\bibinfo {title} {Long-time behavior of isolated periodically driven interacting lattice systems},\ }\href {https://doi.org/10.1103/PhysRevX.4.041048} {\bibfield  {journal} {\bibinfo  {journal} {Phys. Rev. X}\ }\textbf {\bibinfo {volume} {4}},\ \bibinfo {pages} {041048} (\bibinfo {year} {2014})}\BibitemShut {NoStop}%
\bibitem [{\citenamefont {Ueda}(2020)}]{Uedathermalization}%
  \BibitemOpen
  \bibfield  {author} {\bibinfo {author} {\bibfnamefont {M.}~\bibnamefont {Ueda}},\ }\bibfield  {title} {\bibinfo {title} {Quantum equilibration, thermalization and prethermalization in ultracold atoms},\ }\href {https://doi.org/10.1038/s42254-020-0237-x} {\bibfield  {journal} {\bibinfo  {journal} {Nature Reviews Physics}\ }\textbf {\bibinfo {volume} {2}},\ \bibinfo {pages} {669} (\bibinfo {year} {2020})}\BibitemShut {NoStop}%
\bibitem [{\citenamefont {Haldar}\ and\ \citenamefont {Das}(2022)}]{statisticalmechanicsoffloquetquantummatterexactandemergentconservationlaws}%
  \BibitemOpen
  \bibfield  {author} {\bibinfo {author} {\bibfnamefont {A.}~\bibnamefont {Haldar}}\ and\ \bibinfo {author} {\bibfnamefont {A.}~\bibnamefont {Das}},\ }\bibfield  {title} {\bibinfo {title} {Statistical mechanics of floquet quantum matter: exact and emergent conservation laws},\ }\href {https://doi.org/10.1088/1361-648x/ac03d2} {\bibfield  {journal} {\bibinfo  {journal} {Journal of Physics: Condensed Matter}\ }\textbf {\bibinfo {volume} {34}},\ \bibinfo {pages} {234001} (\bibinfo {year} {2022})}\BibitemShut {NoStop}%
\bibitem [{\citenamefont {Haldar}\ \emph {et~al.}(2018)\citenamefont {Haldar}, \citenamefont {Moessner},\ and\ \citenamefont {Das}}]{manybodyonsetoffloquetthermalization}%
  \BibitemOpen
  \bibfield  {author} {\bibinfo {author} {\bibfnamefont {A.}~\bibnamefont {Haldar}}, \bibinfo {author} {\bibfnamefont {R.}~\bibnamefont {Moessner}},\ and\ \bibinfo {author} {\bibfnamefont {A.}~\bibnamefont {Das}},\ }\bibfield  {title} {\bibinfo {title} {Onset of floquet thermalization},\ }\bibfield  {journal} {\bibinfo  {journal} {Physical Review B}\ }\textbf {\bibinfo {volume} {97}},\ \href {https://doi.org/10.1103/physrevb.97.245122} {10.1103/physrevb.97.245122} (\bibinfo {year} {2018})\BibitemShut {NoStop}%
\bibitem [{\citenamefont {Russomanno}\ \emph {et~al.}(2012)\citenamefont {Russomanno}, \citenamefont {Silva},\ and\ \citenamefont {Santoro}}]{periodicsteadyregimeandinterferenceinaperiodicallydrivenquantumsystem}%
  \BibitemOpen
  \bibfield  {author} {\bibinfo {author} {\bibfnamefont {A.}~\bibnamefont {Russomanno}}, \bibinfo {author} {\bibfnamefont {A.}~\bibnamefont {Silva}},\ and\ \bibinfo {author} {\bibfnamefont {G.~E.}\ \bibnamefont {Santoro}},\ }\bibfield  {title} {\bibinfo {title} {Periodic steady regime and interference in a periodically driven quantum system},\ }\href {https://doi.org/10.1103/PhysRevLett.109.257201} {\bibfield  {journal} {\bibinfo  {journal} {Phys. Rev. Lett.}\ }\textbf {\bibinfo {volume} {109}},\ \bibinfo {pages} {257201} (\bibinfo {year} {2012})}\BibitemShut {NoStop}%
\bibitem [{\citenamefont {Seetharam}\ \emph {et~al.}(2018)\citenamefont {Seetharam}, \citenamefont {Titum}, \citenamefont {Kolodrubetz},\ and\ \citenamefont {Refael}}]{absenceofthermalizationinfiniteisolatedinteractingfloquetsystems1}%
  \BibitemOpen
  \bibfield  {author} {\bibinfo {author} {\bibfnamefont {K.}~\bibnamefont {Seetharam}}, \bibinfo {author} {\bibfnamefont {P.}~\bibnamefont {Titum}}, \bibinfo {author} {\bibfnamefont {M.}~\bibnamefont {Kolodrubetz}},\ and\ \bibinfo {author} {\bibfnamefont {G.}~\bibnamefont {Refael}},\ }\bibfield  {title} {\bibinfo {title} {Absence of thermalization in finite isolated interacting floquet systems},\ }\href {https://doi.org/10.1103/PhysRevB.97.014311} {\bibfield  {journal} {\bibinfo  {journal} {Phys. Rev. B}\ }\textbf {\bibinfo {volume} {97}},\ \bibinfo {pages} {014311} (\bibinfo {year} {2018})}\BibitemShut {NoStop}%
\bibitem [{\citenamefont {Bukov}\ and\ \citenamefont {Heyl}(2012)}]{parametricinstabilityinperiodicallydrivenluttingerliquids}%
  \BibitemOpen
  \bibfield  {author} {\bibinfo {author} {\bibfnamefont {M.}~\bibnamefont {Bukov}}\ and\ \bibinfo {author} {\bibfnamefont {M.}~\bibnamefont {Heyl}},\ }\bibfield  {title} {\bibinfo {title} {Parametric instability in periodically driven luttinger liquids},\ }\bibfield  {journal} {\bibinfo  {journal} {Physical Review B}\ }\textbf {\bibinfo {volume} {86}},\ \href {https://doi.org/10.1103/physrevb.86.054304} {10.1103/physrevb.86.054304} (\bibinfo {year} {2012})\BibitemShut {NoStop}%
\bibitem [{\citenamefont {Zhang}\ \emph {et~al.}(2017)\citenamefont {Zhang}, \citenamefont {Hess}, \citenamefont {Kyprianidis}, \citenamefont {Becker}, \citenamefont {Lee}, \citenamefont {Smith}, \citenamefont {Pagano}, \citenamefont {Potirniche}, \citenamefont {Potter}, \citenamefont {Vishwanath}, \citenamefont {Yao},\ and\ \citenamefont {Monroe}}]{timecrystal}%
  \BibitemOpen
  \bibfield  {author} {\bibinfo {author} {\bibfnamefont {J.}~\bibnamefont {Zhang}}, \bibinfo {author} {\bibfnamefont {P.~W.}\ \bibnamefont {Hess}}, \bibinfo {author} {\bibfnamefont {A.}~\bibnamefont {Kyprianidis}}, \bibinfo {author} {\bibfnamefont {P.}~\bibnamefont {Becker}}, \bibinfo {author} {\bibfnamefont {A.}~\bibnamefont {Lee}}, \bibinfo {author} {\bibfnamefont {J.}~\bibnamefont {Smith}}, \bibinfo {author} {\bibfnamefont {G.}~\bibnamefont {Pagano}}, \bibinfo {author} {\bibfnamefont {I.-D.}\ \bibnamefont {Potirniche}}, \bibinfo {author} {\bibfnamefont {A.~C.}\ \bibnamefont {Potter}}, \bibinfo {author} {\bibfnamefont {A.}~\bibnamefont {Vishwanath}}, \bibinfo {author} {\bibfnamefont {N.~Y.}\ \bibnamefont {Yao}},\ and\ \bibinfo {author} {\bibfnamefont {C.}~\bibnamefont {Monroe}},\ }\bibfield  {title} {\bibinfo {title} {Observation of a discrete time crystal},\ }\href {https://doi.org/10.1038/nature21413} {\bibfield  {journal} {\bibinfo  {journal} {Nature}\ }\textbf {\bibinfo {volume} {543}},\ \bibinfo
  {pages} {217–220} (\bibinfo {year} {2017})}\BibitemShut {NoStop}%
\bibitem [{\citenamefont {Khemani}\ \emph {et~al.}(2019)\citenamefont {Khemani}, \citenamefont {Moessner},\ and\ \citenamefont {Sondhi}}]{khemani2019briefhistorytimecrystals}%
  \BibitemOpen
  \bibfield  {author} {\bibinfo {author} {\bibfnamefont {V.}~\bibnamefont {Khemani}}, \bibinfo {author} {\bibfnamefont {R.}~\bibnamefont {Moessner}},\ and\ \bibinfo {author} {\bibfnamefont {S.~L.}\ \bibnamefont {Sondhi}},\ }\href {https://arxiv.org/abs/1910.10745} {\bibinfo {title} {A brief history of time crystals}} (\bibinfo {year} {2019}),\ \Eprint {https://arxiv.org/abs/1910.10745} {arXiv:1910.10745 [cond-mat.str-el]} \BibitemShut {NoStop}%
\bibitem [{\citenamefont {Na}\ \emph {et~al.}(2024)\citenamefont {Na}, \citenamefont {Kemp}, \citenamefont {Griffin},\ and\ \citenamefont {Peng}}]{na2024engineeringmicromotionfloquetprethermalization}%
  \BibitemOpen
  \bibfield  {author} {\bibinfo {author} {\bibfnamefont {I.}~\bibnamefont {Na}}, \bibinfo {author} {\bibfnamefont {J.}~\bibnamefont {Kemp}}, \bibinfo {author} {\bibfnamefont {S.~M.}\ \bibnamefont {Griffin}},\ and\ \bibinfo {author} {\bibfnamefont {Y.}~\bibnamefont {Peng}},\ }\href {https://arxiv.org/abs/2412.09577} {\bibinfo {title} {Engineering micromotion in floquet prethermalization via space-time symmetries}} (\bibinfo {year} {2024}),\ \Eprint {https://arxiv.org/abs/2412.09577} {arXiv:2412.09577 [quant-ph]} \BibitemShut {NoStop}%
\bibitem [{\citenamefont {Xu}\ and\ \citenamefont {Wu}(2018)}]{spacetimecrystal}%
  \BibitemOpen
  \bibfield  {author} {\bibinfo {author} {\bibfnamefont {S.}~\bibnamefont {Xu}}\ and\ \bibinfo {author} {\bibfnamefont {C.}~\bibnamefont {Wu}},\ }\bibfield  {title} {\bibinfo {title} {Space-time crystal and space-time group},\ }\href {https://doi.org/10.1103/PhysRevLett.120.096401} {\bibfield  {journal} {\bibinfo  {journal} {Phys. Rev. Lett.}\ }\textbf {\bibinfo {volume} {120}},\ \bibinfo {pages} {096401} (\bibinfo {year} {2018})}\BibitemShut {NoStop}%
\bibitem [{\citenamefont {Aubry}\ and\ \citenamefont {André}(1980)}]{articleaah}%
  \BibitemOpen
  \bibfield  {author} {\bibinfo {author} {\bibfnamefont {S.}~\bibnamefont {Aubry}}\ and\ \bibinfo {author} {\bibfnamefont {G.}~\bibnamefont {André}},\ }\bibfield  {title} {\bibinfo {title} {Analyticity breaking and anderson localization in incommensurate lattices},\ }\href@noop {} {\bibfield  {journal} {\bibinfo  {journal} {Proceedings, VIII International Colloquium on Group-Theoretical Methods in Physics}\ }\textbf {\bibinfo {volume} {3}} (\bibinfo {year} {1980})}\BibitemShut {NoStop}%
\bibitem [{\citenamefont {Abualnaja}\ \emph {et~al.}(2023)\citenamefont {Abualnaja}, \citenamefont {He}, \citenamefont {Andreev}, \citenamefont {Jones},\ and\ \citenamefont {Durrani}}]{singleparticleentropy}%
  \BibitemOpen
  \bibfield  {author} {\bibinfo {author} {\bibfnamefont {F.}~\bibnamefont {Abualnaja}}, \bibinfo {author} {\bibfnamefont {W.}~\bibnamefont {He}}, \bibinfo {author} {\bibfnamefont {A.}~\bibnamefont {Andreev}}, \bibinfo {author} {\bibfnamefont {M.}~\bibnamefont {Jones}},\ and\ \bibinfo {author} {\bibfnamefont {Z.}~\bibnamefont {Durrani}},\ }\bibfield  {title} {\bibinfo {title} {Single particle entropy stability and the temperature-entropy diagram in quantum dot transistors},\ }\href {https://doi.org/10.1103/PhysRevResearch.5.033025} {\bibfield  {journal} {\bibinfo  {journal} {Phys. Rev. Res.}\ }\textbf {\bibinfo {volume} {5}},\ \bibinfo {pages} {033025} (\bibinfo {year} {2023})}\BibitemShut {NoStop}%
\bibitem [{\citenamefont {Russomanno}\ and\ \citenamefont {Santoro}(2017)}]{floquetresonancesclosetotheadiabaticlimit}%
  \BibitemOpen
  \bibfield  {author} {\bibinfo {author} {\bibfnamefont {A.}~\bibnamefont {Russomanno}}\ and\ \bibinfo {author} {\bibfnamefont {G.~E.}\ \bibnamefont {Santoro}},\ }\bibfield  {title} {\bibinfo {title} {Floquet resonances close to the adiabatic limit and the effect of dissipation},\ }\href {https://doi.org/10.1088/1742-5468/aa8702} {\bibfield  {journal} {\bibinfo  {journal} {Journal of Statistical Mechanics: Theory and Experiment}\ }\textbf {\bibinfo {volume} {2017}},\ \bibinfo {pages} {103104} (\bibinfo {year} {2017})}\BibitemShut {NoStop}%
\bibitem [{\citenamefont {Hu}\ \emph {et~al.}(2023)\citenamefont {Hu}, \citenamefont {Fu}, \citenamefont {Li},\ and\ \citenamefont {Shen}}]{solvablemodelfordiscretetimecrystalenforcedbynonsymmorphicdynamicalsymmetry}%
  \BibitemOpen
  \bibfield  {author} {\bibinfo {author} {\bibfnamefont {Z.-A.}\ \bibnamefont {Hu}}, \bibinfo {author} {\bibfnamefont {B.}~\bibnamefont {Fu}}, \bibinfo {author} {\bibfnamefont {X.}~\bibnamefont {Li}},\ and\ \bibinfo {author} {\bibfnamefont {S.-Q.}\ \bibnamefont {Shen}},\ }\bibfield  {title} {\bibinfo {title} {Solvable model for discrete time crystal enforced by nonsymmorphic dynamical symmetry},\ }\href {https://doi.org/10.1103/PhysRevResearch.5.L032024} {\bibfield  {journal} {\bibinfo  {journal} {Phys. Rev. Res.}\ }\textbf {\bibinfo {volume} {5}},\ \bibinfo {pages} {L032024} (\bibinfo {year} {2023})}\BibitemShut {NoStop}%
\bibitem [{\citenamefont {Hone}\ \emph {et~al.}(1997)\citenamefont {Hone}, \citenamefont {Ketzmerick},\ and\ \citenamefont {Kohn}}]{timedependentfloquettheoryandabsenceofanadiabaticlimit}%
  \BibitemOpen
  \bibfield  {author} {\bibinfo {author} {\bibfnamefont {D.~W.}\ \bibnamefont {Hone}}, \bibinfo {author} {\bibfnamefont {R.}~\bibnamefont {Ketzmerick}},\ and\ \bibinfo {author} {\bibfnamefont {W.}~\bibnamefont {Kohn}},\ }\bibfield  {title} {\bibinfo {title} {Time-dependent floquet theory and absence of an adiabatic limit},\ }\href {https://doi.org/10.1103/PhysRevA.56.4045} {\bibfield  {journal} {\bibinfo  {journal} {Phys. Rev. A}\ }\textbf {\bibinfo {volume} {56}},\ \bibinfo {pages} {4045} (\bibinfo {year} {1997})}\BibitemShut {NoStop}%
\bibitem [{\citenamefont {Ishii}\ \emph {et~al.}(2018)\citenamefont {Ishii}, \citenamefont {Kuwahara}, \citenamefont {Mori},\ and\ \citenamefont {Hatano}}]{heatinginintegrabletimeperiodicsystems}%
  \BibitemOpen
  \bibfield  {author} {\bibinfo {author} {\bibfnamefont {T.}~\bibnamefont {Ishii}}, \bibinfo {author} {\bibfnamefont {T.}~\bibnamefont {Kuwahara}}, \bibinfo {author} {\bibfnamefont {T.}~\bibnamefont {Mori}},\ and\ \bibinfo {author} {\bibfnamefont {N.}~\bibnamefont {Hatano}},\ }\bibfield  {title} {\bibinfo {title} {Heating in integrable time-periodic systems},\ }\href {https://doi.org/10.1103/PhysRevLett.120.220602} {\bibfield  {journal} {\bibinfo  {journal} {Phys. Rev. Lett.}\ }\textbf {\bibinfo {volume} {120}},\ \bibinfo {pages} {220602} (\bibinfo {year} {2018})}\BibitemShut {NoStop}%
\bibitem [{\citenamefont {Eckardt}\ and\ \citenamefont {Anisimovas}(2015)}]{perturbativefloquethighfrequency}%
  \BibitemOpen
  \bibfield  {author} {\bibinfo {author} {\bibfnamefont {A.}~\bibnamefont {Eckardt}}\ and\ \bibinfo {author} {\bibfnamefont {E.}~\bibnamefont {Anisimovas}},\ }\bibfield  {title} {\bibinfo {title} {High-frequency approximation for periodically driven quantum systems from a floquet-space perspective},\ }\href {https://doi.org/10.1088/1367-2630/17/9/093039} {\bibfield  {journal} {\bibinfo  {journal} {New Journal of Physics}\ }\textbf {\bibinfo {volume} {17}},\ \bibinfo {pages} {093039} (\bibinfo {year} {2015})}\BibitemShut {NoStop}%
\bibitem [{\citenamefont {Rodriguez-Vega}\ \emph {et~al.}(2018)\citenamefont {Rodriguez-Vega}, \citenamefont {Lentz},\ and\ \citenamefont {Seradjeh}}]{floquetperturbationtheory}%
  \BibitemOpen
  \bibfield  {author} {\bibinfo {author} {\bibfnamefont {M.}~\bibnamefont {Rodriguez-Vega}}, \bibinfo {author} {\bibfnamefont {M.}~\bibnamefont {Lentz}},\ and\ \bibinfo {author} {\bibfnamefont {B.}~\bibnamefont {Seradjeh}},\ }\bibfield  {title} {\bibinfo {title} {Floquet perturbation theory: formalism and application to low-frequency limit},\ }\href {https://doi.org/10.1088/1367-2630/aade37} {\bibfield  {journal} {\bibinfo  {journal} {New Journal of Physics}\ }\textbf {\bibinfo {volume} {20}},\ \bibinfo {pages} {093022} (\bibinfo {year} {2018})}\BibitemShut {NoStop}%
\bibitem [{\citenamefont {Hewitt}\ and\ \citenamefont {Hewitt}(1979)}]{Gibbsphenomenon}%
  \BibitemOpen
  \bibfield  {author} {\bibinfo {author} {\bibfnamefont {E.}~\bibnamefont {Hewitt}}\ and\ \bibinfo {author} {\bibfnamefont {R.~E.}\ \bibnamefont {Hewitt}},\ }\bibfield  {title} {\bibinfo {title} {The {Gibbs-Wilbraham} phenomenon: An episode in fourier analysis},\ }\href {https://doi.org/10.1007/BF00330404} {\bibfield  {journal} {\bibinfo  {journal} {Archive for History of Exact Sciences}\ }\textbf {\bibinfo {volume} {21}},\ \bibinfo {pages} {129} (\bibinfo {year} {1979})}\BibitemShut {NoStop}%
\bibitem [{\citenamefont {Chandran}\ \emph {et~al.}(2023)\citenamefont {Chandran}, \citenamefont {Iadecola}, \citenamefont {Khemani},\ and\ \citenamefont {Moessner}}]{quantummanybodyscars}%
  \BibitemOpen
  \bibfield  {author} {\bibinfo {author} {\bibfnamefont {A.}~\bibnamefont {Chandran}}, \bibinfo {author} {\bibfnamefont {T.}~\bibnamefont {Iadecola}}, \bibinfo {author} {\bibfnamefont {V.}~\bibnamefont {Khemani}},\ and\ \bibinfo {author} {\bibfnamefont {R.}~\bibnamefont {Moessner}},\ }\bibfield  {title} {\bibinfo {title} {Quantum many-body scars: A quasiparticle perspective},\ }\href {https://doi.org/https://doi.org/10.1146/annurev-conmatphys-031620-101617} {\bibfield  {journal} {\bibinfo  {journal} {Annual Review of Condensed Matter Physics}\ }\textbf {\bibinfo {volume} {14}},\ \bibinfo {pages} {443} (\bibinfo {year} {2023})}\BibitemShut {NoStop}%
\end{thebibliography}%

\end{document}


\vspace*{0.2cm}
\noindent\textbf{Supplemental Material for:} \\ 
\noindent\textbf{``Spontaneous formation of subsystem and bath under accordion-type driving''} \\ 
\hrule
\vspace{0.5cm}
\section{Additional numerical results}
\subsection{Finite Size Effect}
\begin{figure}[htbp]
    \centering
    \includegraphics[width=0.6\linewidth]{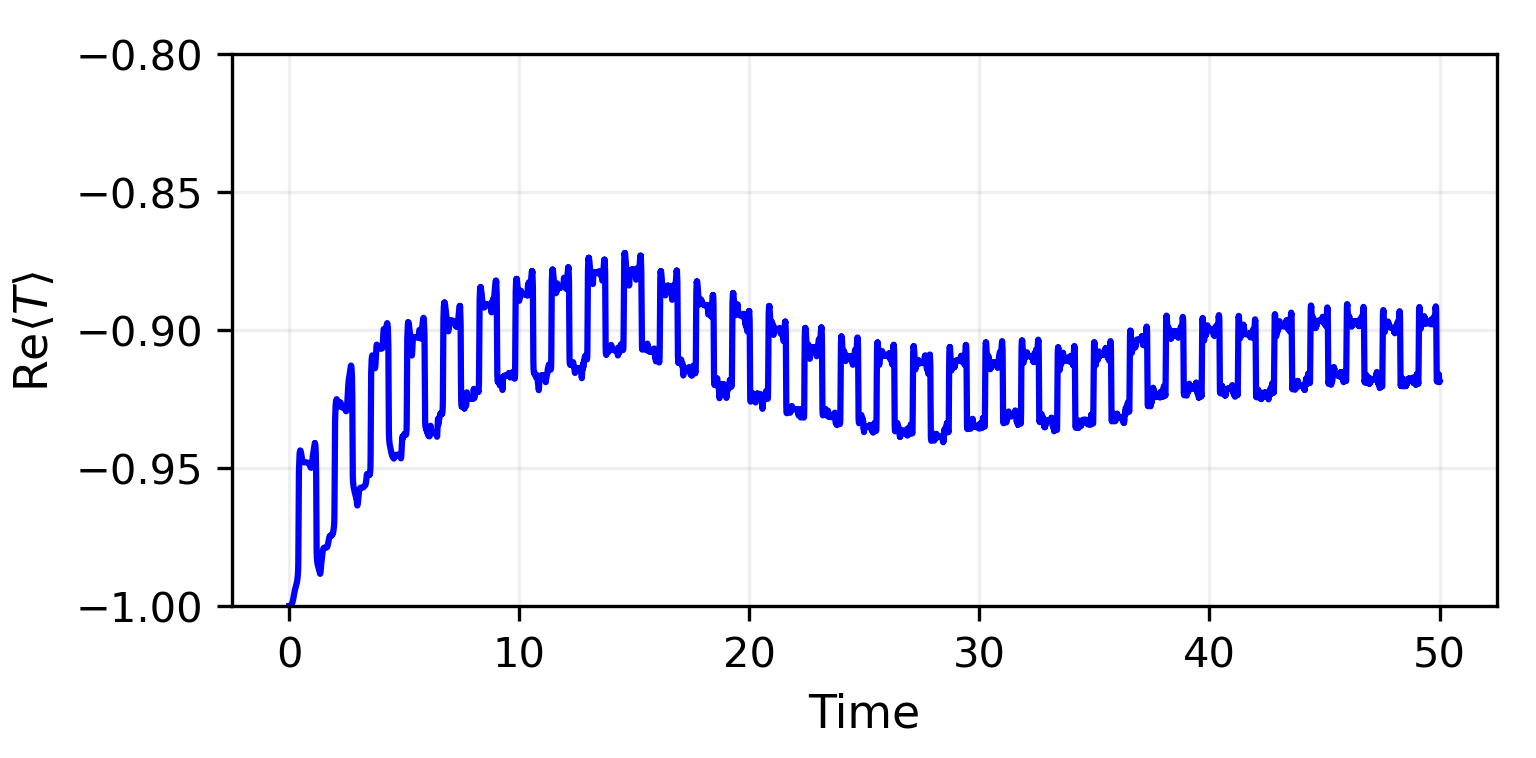}
    \caption{ The evolution of the translational operator with a initial momentum $k=\pi$ in $\omega=4,V=5,h=-1$. However this size is $N=800$, rather than $N=200$ in the main paper
    }
    \label{fig:sample}
\end{figure}

As the onsite potential is a Bessel function which decays slowly with respect to length and there are incommensurate phenomenon at some time, the finite size effect is necessary to check in the numerical result. We test the dynamical structure from $N=200$ to $N=800$ as shown in Fig.~\ref{fig:sample}.

The square wave is clearer, which means that our numerical result does not come from an accidental finite size effect. More importantly, the large size of $N$ can suppress the Gibbs phenomenon.

\subsection{Shape of wavefunction}
\begin{figure}[htbp]
    \centering
    \begin{minipage}{0.48\textwidth}
        \centering
        \includegraphics[width=\linewidth]{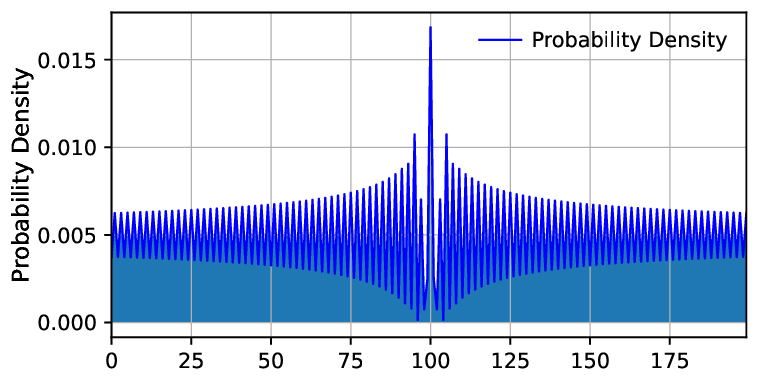}
        \caption{Probability distribution of wavefunction in the subsystem with square wave condition.}
        \label{supfig-fig2b}
    \end{minipage}
    \hfill
    \begin{minipage}{0.48\textwidth}
        \centering
        \includegraphics[width=\linewidth]{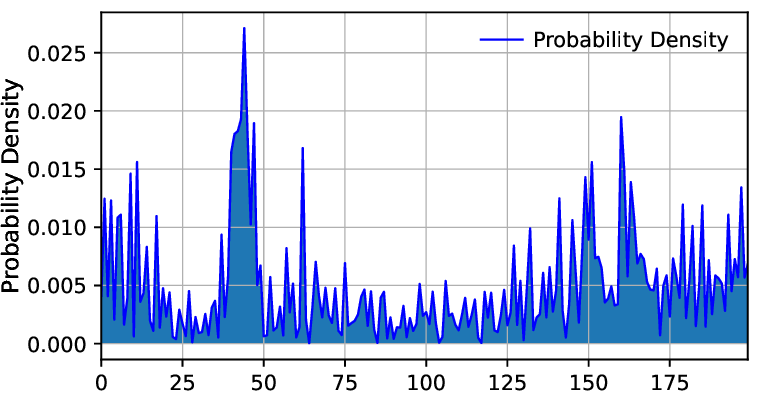}
        \caption{Random wavefunction distribution when not in the subsystem.}
        \label{supfig-fig3b}
    \end{minipage}
\end{figure}

To show the characters of the evolution directly, we show two typical wavefunctions of the evolution in Fig.~\ref{supfig-fig2b} and Fig.~\ref{supfig-fig3b}.

\section{Theoretical Derivations}
\subsection{Single particle entanglement entropy}

Usually, the entanglement entropy can only be defined in many-body problem to understand the thermalization. This is because the Hilbert space of the many-body Hamiltonian is described by a structure of tensor product with dimension $D=d^{L}$ so that the total Hilbert space can be separated to the tensor product of two parts $d^{L_{A}+L_{B}}=d^{L_{A}}\times d^{L_{B}}$, which is subsystem and environment, but this structure will become trivial when $d=1$, where $d$ is the size of quantum states at one site and $L$ is the size of the lattice.

However, the entanglement can be generalized to different freedoms, and there is no reason that we cannot define the entanglement entropy of the single-particle wavefunction on a lattice system to describe the entanglement in space. There is indeed some attempt to interpret this as Shannon entropy\cite{singleparticleentropy}, which is defined as
\begin{equation}
    S=-p_{A}\ln(p_{A})-p_{B}\ln(p_{B}),p_{A}+p_{B}=1,
\end{equation}
where $p_{A}=\sum_{i=1}^{N_{A}}|\psi_{i}|^{2}$, but this definition is a classical interpretation that regards the wavefunction as a particle's probability to fall into one of the two boxes. A natural generalization should cover the properties of related phase between the two parts. 
Thus, we propose a new quantity to describe the entanglement entropy of this bipartite system. 
\begin{equation}
     S_{en}(t)=-\mathrm{Tr}[\rho\ln(\rho)],\rho=\ket{\psi_{A}}\bra{\psi_{A}}+\ket{\psi_{B}}\bra{\psi_{B}},
 \end{equation}
 where $\psi_{A}$ is the part supported on set $A=L/2$ of total wavefunction without normalization. On the basis of $\{\ket{\psi_{A}},\ket{\psi_{B}}\}$, there are actually these equalities
 \begin{equation}
    \begin{split}
&\bra{\psi_{A}}\ket{\psi_{A}}=p_{A},\bra{\psi_{B}}\ket{\psi_{B}}=p_{B},\bra{\psi_{B}}\ket{\psi_{A}}=c,\\
&\rho=
\begin{bmatrix}
    p_{A}^{2}+c^{2} & c^{\star}\\
    c & p_{B}^{2}+c^{2}
\end{bmatrix}.
    \end{split}
 \end{equation}
 The problem of this definition is that $\rho$ here is not a density matrix which satisfies $\text{Tr}[\rho]=1$. However, this quantity actually covers the random properties of probability distribution and related phase simultaneously.
 
 We show the properties of this entropy as following, if the two states are parallel with $\ket{\psi_{B}}=m\ket{\psi_{A}}$, the probability restriction will be $(m^{2}+1)p_{A}=1$ so that $\rho=(m^{2}+1)\ket{\psi_{A}}\bra{\psi_{A}}=\frac{1}{p_{A}}\ket{\psi_{A}}\bra{\psi_{A}}$ and it has only one eigenvalue $\lambda=1$, which means that the entropy will be zero. If the two states are orthogonal, the entropy will be
 \begin{equation}
     S=-2p_{A}^{2}\ln(p_{A})-2p_{B}^{2}\ln (p_{B}),
 \end{equation}
 and when $p_{A}=p_{B}=0.5$, it will achieve its maximal quantity $\ln 2$.
 More rigorously, the eigenvalues of $\rho$ satisfy
 \begin{equation}
     \lambda^{2}-(2c^{2}+p_{A}^{2}+p_{B}^{2})\lambda +c^{4}+c^{2}(p_{A}^{2}+p_{B}^{2})+p_{A}^{2}p_{B}^{2}-c^{2}=0,
     \label{eq-eigenequation}
 \end{equation}
and the entropy is 
\begin{equation}
    S=-\lambda_{1}\ln\lambda_{1}-\lambda_{2} \ln \lambda_{2},
\end{equation}
where $\lambda_{1},\lambda_{2}$ are the two eigenvalues of Eq.~\ref{eq-eigenequation}.
With Cauchy-Schwarz inequality, we get the range of $c$
\begin{equation}
    c^{2} \leq p_{A}^{2}p_{B}^{2}\leq \frac{1}{16}.
\end{equation}
Then, it can be proven that 
\begin{equation}
    \lambda_{1}\lambda_{2}=c^{4}+c^{2}(p_{A}^{2}+p_{B}^{2})+p_{A}^{2}p_{B}^{2}-c^{2}\geq 0,
\end{equation}
with $p_{A}+p_{B}=1,c^{2}\in[0,\frac{1}{16}]$ and so the entropy is well defined, which can also be explained as $\bra{\phi}\rho \ket{\phi}>0$ for any $\ket{\phi}$. More importantly, it can be checked numerically that among all the values of $c^{2}$, the entropy
\begin{equation}
S(c^{2}) = \max_{p_{A} \in [0,1]} \bigl( -\lambda_{1} \ln \lambda_{1} - \lambda_{2} \ln \lambda_{2} \bigr)  
\end{equation}
will decrease with $c^{2}$, which means that only $c=0$ can obtain the maximal entropy $\ln2$.

 In summary, this quantity contains the behavior of the Shannon entropy where the most random situation is achieved when $p_{A}=0.5$. More importantly, it also contains the property of quantum entanglement described by the randomness of the related phase at different sites that $\bra{\psi_{B}}\ket{\psi_{A}}=0$.
\subsection{Floquet theory, Magnus expansion and Fourier transformation of Bessel function}

In this part, we introduce the method of Floquet Magnus expansion and derive the effective Hamiltonian of our accordion model in the momentum space which is divergent in the 2nd order. 

Floquet theory based on the operators is intended to decompose the evolution operator as a macro-motion on the $t=nT$ lattice and a micro-motion in a period\cite{Bukov_review}. The typical picture of this is to make a time-dependent rotational transformation to a handlable Hamiltonian and then rotate back, which is so general that it can be applied to handle some important static models like Anderson model which is known as Schrieffer-Wolff transformation. Most discussions of this method are about the effective description of the macro-motion, whose Floquet gauge about the initial time should be discussed in detail. Here, among the many methods, we introduce the Floquet-Magnus expansion, which is independent of the Floquet gauge. 

We want to know the evolution operator of the time-dependent Hamiltonian which is defined as:
\begin{equation}
    U(t_{2},t_{1})=T_{t}\exp[-i\int_{t_{1}}^{t_{2}}H(\tilde{t})d\tilde{t}]=e^{-iK_{\text{F}}[t_{0}](t_{2})}e^{-iH_{\text{F}}[t_{0}](t_{2}-t_{1})}e^{iK_{\text{F}}[t_{0}](t_{1})}.
\end{equation}
where $K_{\text{F}}$ is the stroboscopic kick operator depending explicitly on the Floquet gauge $t_{0}$ which is defined as:
\begin{equation}
    e^{-iK_{\text{F}}[t_{0}](t_{2})}=U(t_{2},t_{0})e^{iH_{\text{F}}[t_{0}](t_{2}-t_{0})}.
\end{equation}
It can be proven that the effective description derived by Floquet Magnus expansion is irrelavent to the Floquet gauge $t_{0}$\cite{Bukov_review}. Thus,
considering the differentiation of this equation, we get 
\begin{equation}
    H_{\text{F}}=e^{iK_{\text{F}}(t)}H(t)e^{-iK_{\text{F}}(t)}-ie^{iK_{\text{F}}(t)}(\frac{d}{dt}e^{-iK_{\text{F}}(t)}).
\end{equation}
Then, we make a perturbative expansion as
\begin{equation}
    H_{\text{F}}=\sum_{n=0}^{\infty}H_{\text{F}}^{(n)},K_{\text{F}}(t)=\sum_{n=0}^{\infty}K_{\text{F}}^{(n)}(t),H(t)=\sum_{l\in Z}H_{l}e^{il\Omega t},
\end{equation}
and we will have a method to get $H_{\text{F}}$ from the Fourier coefficients of the $H(t)$
\begin{equation}
    \begin{split}
     H_{\text{F}}^{(0)}&=H_{0},K_{\text{F}}^{(1)}=\sum_{l\neq 0}\frac{e^{il\omega t}}{il\omega}H_{l}, \\
     H_{\text{F}}^{(1)}&= \sum_{l\neq 0}\frac{1}{2l\omega}[H_{l},H_{-l}],\\H_{\text{F}}^{(2)}&=\sum_{l\neq 0}\frac{1}{2l^{2}\omega^{2}}[[H_{l},H_{0}],H_{-l}]+\sum_{l,l',l+l'\neq 0}\frac{1}{3ll'\omega^{2}}[H_{l},[H_{l'},H_{-l-l'}]],
    \end{split}
\end{equation}
which are the necessary results for our model here. 

Next, we come back to our model
\begin{equation}
    H=\sum_{n}\{\sum_{i}J_{2n}(\pi  i)e^{i2n\omega t}c_{i}^{\dagger}c_{i}+\delta_{n,0}hc_{i+1}^{\dagger}c_{i}+\text{h.c}\}.
\end{equation}
As the only term which can contribute to the commutators is $H_{0}$ with the hopping term in the lattice sites, the results of expansion is  
\begin{equation}
  \begin{split}
   H^{(0)}_{\text{F}}&=\sum_{i}VJ_{0}(\pi i)c_{i}^{\dagger}c_{i}+hc_{i+1}^{\dagger}c_{i}+\mathrm{h.c.},\\
  H_{\text{F}}^{(1)}&=0, \\
   H^{(2)}_{\text{F}} &=tV^{2}\sum_{m\neq 0}\sum_{i}\frac{(J_{2m}(\pi(i+1))-J_{2m}(\pi i))^{2}}{8m^{2}\omega^{2}}(c^{\dagger}_{i+1}c_{i}+\mathrm{h.c.}).
  \end{split}
\end{equation}
Then we change to momentum space.
This transformation is tractable from the integral representation of Bessel function:
\begin{equation}
    J_{n}(\pi j)=\frac{1}{2\pi}\int_{-\pi}^{\pi}e^{i(n\theta-\pi j \sin\theta)}d\theta,
\end{equation}
where $j$ is the lattice number. We make the discrete Fourier transformation of space, restrict the scattering momentum in the range $q\in (-\pi,\pi]$ and $L\to \infty$, we get
\begin{equation}
    \tilde{J_{n}}(q)= \int_{-\pi}^{\pi}e^{in\theta}\sum_{m=-\infty}^{\infty}\delta(q+\pi \sin \theta-2\pi m)d\theta=e^{-in\arcsin(q/\pi)}\frac{1}{\pi\sqrt{1-(q/\pi)^{2}}},q\in(-\pi,\pi],
\end{equation}
where the Poisson summation formula is used. When considering the subsystem with scattering in singularity $q= \pi$, the phase factor will be $e^{- in\frac{\pi}{2}}$.

With these properties, we can analyze the Magnus expansion in the momentum space.
\begin{equation}
    H_ {\text{F}}^{(0)}(k)=2h\cos(k)c_{k}^{\dagger}c_{k}+\sum_{q}\frac{V(c_{k+q}^{\dagger}c_{k}+\mathrm{h.c.})}{\pi\sqrt{1-({q / \pi}})^{2}},
\end{equation}
where $q$ is the scattering momentum, and the momentum $k$ is restricted to the first Brillouin zone$[0,2\pi)$.
Meanwhile, denoting $F(i)=\frac{(J_{2m}(\pi(i+1))-J_{2m}(\pi i))^{2}}{8m^{2}\omega^{2}}$, we can get \begin{equation}
    H^{(2)}_{\mathrm F}
= tV^2\frac{1}{N}\sum_{k,q}\widetilde F_{\,q-k}\;e^{-ik}\;c_k^\dagger c_q 
\;+\;\text{h.c.}.
\end{equation} The Fourier transformation of $F(i)$ is 
\begin{equation}
\widetilde F_q
= \sum_{m\ne0}\frac{1}{8 m^{2}\omega^{2}}\;\frac{1}{N}\sum_{p}
\big(e^{-ip}-1\big)\big(e^{-i(q-p)}-1\big)\;
\widetilde J_{2m}(p)\,\widetilde J_{2m}(q-p).
\end{equation} 
It can be found that
$\int_{-\pi}^{\pi}(e^{-ip}-1)(e^{-i(q-p)}-1)\tilde{J}_{m}(p)\tilde{J}_{m}(q-p)dp$ is always divergent for any $q,m$. The divergence in Magnus series is claimed to be relevant to chaotic phenomenon in the Floquet system \cite{Kuwahara_2016floquetmagnustheory}, this is consistent with our numerical results in low frequency.  
\subsection{Perturbative result of the time-dependent eigen wavefunction}
In this part we will first introduce another method based on the analogy of the Bloch theorem, which describes systems with discrete spatial translational symmetry \cite{spacetimecrystal,perturbativefloquethighfrequency,floquetperturbationtheory} and will focus on the eigen wavefunction rather than operators. Then, we derive the results of the perturbative analysis in the subsystem of our accordion model.

Based on the Floquet theorem, the evolution of wavefunction comes from the form
\begin{equation}
    \ket{\psi(t)}=\sum_{l}a_{l}e^{-i\epsilon_{l}t}\ket{\psi_{l}(t)},
\end{equation}
where $\ket{\psi_{l}(t)}$ is an time-dependent eigen wavefunction with quasi energy $\epsilon_{l}$ and $\ket{\psi_{l}(t)}$ is a periodic function of t with frequency $\omega$. So we have
\begin{equation}
    \ket{\psi_{l}}=\sum_{n}e^{-in\omega t}\ket{\phi_{n}}=\sum_{n\alpha}c_{n\alpha}e^{-in\omega t}\ket{k_{\alpha}},
    \label{eq-eigenmodeexpansion}
\end{equation}
where the Hilbert space can be regarded as $\{e^{-in\omega t}\otimes \ket{k_{\alpha}}\}$.
The next step is to get the eigen problem on this basis which can be understood as coupling with different photons with frequency $n\omega$, which is 
\begin{equation}
\begin{split}
    &\sum_{n\beta}H_{m\alpha,n\beta}c_{n \beta}=\epsilon_{l}c_{m\alpha},\\
    &H_{m\alpha,n\beta}=\bra{k_{\alpha}}H_{m-n}-m\omega \delta_{m,n}\ket{k_{\beta}},
\end{split}
\label{eq-floqueteigen}
\end{equation}
where $c_{n\alpha}$ is the eigenvector.
After solving the eigen problem, we consider $t=0$ to get the coefficients $a_{l}$ and achieve the behavior of the evolution. In this method, the advantages are the specific physical meaning about the scattering with different photons and that the scattering structure can give a perturbative description to the corrections of dynamical structure dependent on time. 

We can understand this advantage from the analysis of our models. Expand the $\ket{\psi_{l}(t)}$ in the space $\{\ket{k}\otimes\ket{n\omega}\}$ where $\ket{n\omega}=e^{-in\omega t}$.
In this basis, the eigen Hamiltonian can be understood as coupling between block matrix as shown in Eq.~\ref{eq-floqueteigen}. The block matrix corresponding the eigen problem of our accordion model is:
\begin{equation}
\mathbf{H} =
\begin{bmatrix}
    \ddots               & \vdots                           & \vdots                               & \vdots               &  \\
    \cdots & \tilde{J_0} + 2\omega\mathbf{I}      & \tilde{J}_{-2}                      & \tilde{J}_{-4}          & \cdots \\
    \cdots & \tilde{J}_{2}            & \tilde{J_0}                      & \tilde{J}_{-2}          & \cdots \\
    \cdots & \tilde{J}_{4}            & \tilde{J}_{2}                   & \tilde{J_0} - 2\omega\mathbf{I} & \cdots \\
                  & \vdots                           & \vdots                               & \vdots               & \ddots
\end{bmatrix},
\end{equation}
where these block matrix take the form:
\begin{equation}
\tilde{J}_{0}=
\begin{bmatrix}
    2h  &  \tilde{V} \\
    \tilde{V} &-2h
\end{bmatrix}
,\tilde{J}_{n}=e^{-in\frac{\pi}{2}}\begin{bmatrix}
    0 & \tilde{V} \\
    \tilde{V} & 0
\end{bmatrix},
\end{equation}
and $\tilde{V}$ is a phenomenological parameter to substitute for the singular scattering. If $\tilde{V}$ is small, we can use the 1-st order perturbative theory directly 
\begin{equation}
    \ket{\psi_{1}(t)}\approx\ket{k=0}-\frac{\tilde{V}}{-4h}\ket{k=\pi}-\sum_{n\neq 0}\frac{\tilde{V}}{-4h+2n\omega}e^{i2n(-\frac{\pi}{2}-\omega t)}\ket{k=\pi}.
\end{equation}
To make the result analytical solvable, we choose $\omega=-4h$. Then the series is
\begin{equation}
    \sum_{n}\frac{\tilde{V}e^{in(-\pi-2\omega t)}}{-4-8n}=\sum_{k}\frac{\tilde{V}e^{i(2k+1)(-\pi/2-\omega t)}}{-4(2k+1)}e^{i\frac{\pi+2\omega t}{2}}=\sum_{k=0}^{\infty}\frac{i\tilde{V}\cos((2k+1)\omega t)}{2(2k+1)}e^{i\frac{\pi+2\omega t}{2}},
\end{equation}
which is actually the Fourier expansion of square wave
\begin{equation}
    \sum_{k=0}^{\infty}\frac{\cos((2k+1)\omega t)}{(2k+1)}=\frac{\pi}{4}\text{sgn}(\cos(\omega t)),
\end{equation}
where
\begin{equation}
    \text{sgn}(x) = 
\begin{cases}
1, & x > 0,\\
0, & x = 0,\\
-1, & x < 0.
\end{cases}
\end{equation}
Thus, the perturbative time-dependent eigen wavefunction will be 
\begin{equation}
\begin{split}
    \ket{\psi_{1}(t)}&\approx\ket{k=0}-\frac{i\pi\tilde{V}}{2\omega}\text{sgn}[\cos(\omega t)]e^{i(\pi/2+\omega t)}
    \ket{k=\pi},\\
    \ket{\psi_{2}(t)} &\approx\ket{k=\pi}-\frac{i\pi\tilde{V}}{2\omega}\text{sgn}[\cos(\omega t)]e^{-i(\pi/2+\omega t)}
    \ket{k=0},
\end{split}
\end{equation}
whose perturbative condition is $\frac{\pi \tilde{V}}{2\omega}\ll 1$. As the time dependence is in the perturbative term, this will not break the normalization of eigen wavefunction.


By this method, we can explain why there is a square wave behavior of expectation value when the initial wavefunction is a plane wave with momentum $k=\pi$. From numerical results, we can confirm that the $\tilde{V}$ is actually small in this case, the evolution is 
\begin{equation}
    \ket{\psi(t)}=a_{1}e^{i2ht}\ket{\psi_{1}(t)}+a_{2}e^{-i2ht}\ket{\psi_{2}(t)}.
\end{equation}
Then the numerical quantity satisfies,
\begin{equation}
    \bra{\psi(t)}\hat{T}\ket{\psi(t)}=a_{1}^{2}\bra{\psi_{1}(t)}\hat{T}\ket{\psi_{1}(t)}+a_{2}^{2}\bra{\psi_{2}(t)}\hat{T}\ket{\psi_{2}(t)}+a_{1}a_{2}^{\star}e^{2iht}\bra{\psi_{2}(t)}\hat{T}\ket{\psi_{1}(t)}+\text{h.c.},
\end{equation}
where the translational operator $\hat{T}$ satisfies 
\begin{equation}
    \hat{T}\ket{k=0}=\ket{k=0},\hat{T}\ket{k=\pi}=-\ket{k=\pi}.
\end{equation}
 As the initial wavefunction is near $\ket{\psi_{2}(t=0)}$, which means $a_{2}\gg a_{1}$, the main part of $\bra{\psi(t)}\hat{T}\ket{\psi(t)}$ is 
 \begin{equation}
     -a_{2}^{2}+a_{1}a_{2}^{\star}e^{i4h t}\frac{i\pi\tilde{V}}{\omega}\text{sgn}[\cos(\omega t)]e^{i(\pi/2+\omega t)}+\text{h.c.}\approx -a_{2}^{2}-(a_{1}a_{2}^{\star}+a_{1}^{\star}a_{2})\frac{\pi\tilde{V}}{\omega}\text{sgn}[\cos(\omega t)],
 \end{equation}
 when $\omega=-4h$.
 This is exactly the result shown in the paper.

\begin{figure}[htbp]
    \centering
    \includegraphics[width=0.6\linewidth]{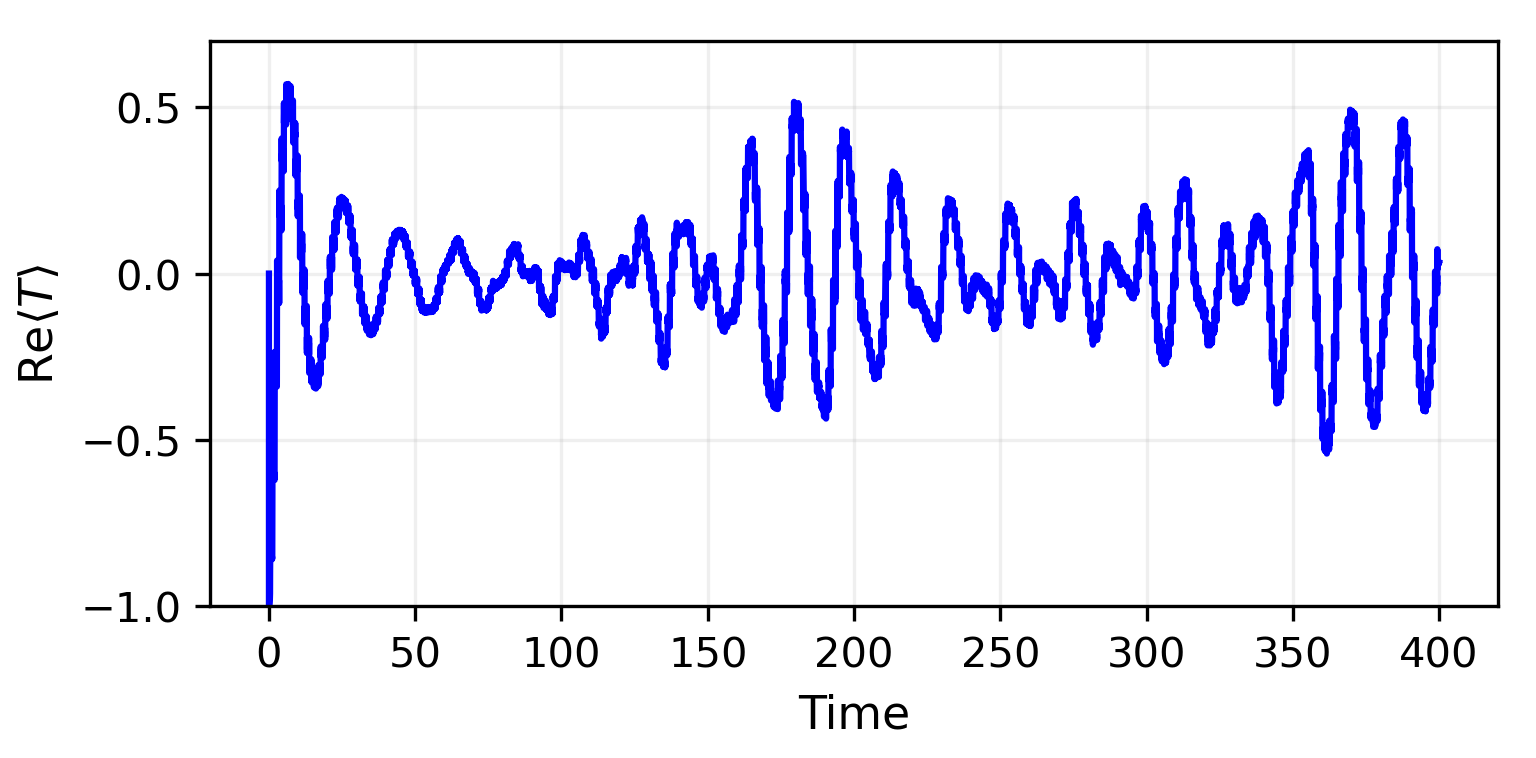}
    \caption{ The evolution of the translational operator with a initial momentum $k=\pi$ in $\omega=4,V=5,h=-2,N=200$.
    }
    \label{fig-resonance}
\end{figure}

Apart from the square wave condition $4h=(2m+1)\omega, m \in \mathbb{Z}$, there is also an important resonant condition that $2h=m\omega, m \in \mathbb{Z}$, which means singularity in the perturbative series. The evolution under this condition will become random, which is defined as random relative phase with the expectation value of $\hat T$ being $0$ as shown in Fig.~\ref{fig-resonance}.
This is because the degenerate perturbative analysis requires a rotation of the Hilbert space, which may rotate the states in the bath back to the subsystem. Interestingly, the problem of incommensurate momentum in AA model is transferred to the incommensurate energy, where commensurate condition $4h=2m\omega$ will cause the failure of perturbative series.     

From a more general perspective, our method of the time-dependent eigen wave function here can be used to analyze many kicked Hamiltonians. For example, if $\tilde{V}$ is large, there is still a method to do the perturbative analysis here. We diagonalize the $\tilde{J}_{0}$ with
\begin{equation}
    V^{\dagger}\tilde{J}_{0}V=\Lambda.
\end{equation}
Then the total unitary transformation is 
\begin{equation}
U=
    \begin{bmatrix}
    \ddots               & \vdots                           & \vdots                               & \vdots               &  \\
    \cdots & V      & 0                      & 0          & \cdots \\
    \cdots & 0            & V                      & 0          & \cdots \\
    \cdots & 0            & 0                   & V & \cdots \\
                  & \vdots                           & \vdots                               & \vdots               & \ddots
\end{bmatrix}.
\end{equation}
Then the Hamiltonian changes to
\begin{equation}
\mathbf{H} =
\begin{bmatrix}
    \ddots               & \vdots                           & \vdots                               & \vdots               &  \\
    \cdots & \Lambda +2 \omega\mathbf{I}      & V^{\dagger}\tilde{J}_{-2}V                      & V^{\dagger}\tilde{J}_{-4}V          & \cdots \\
    \cdots & V^{\dagger}\tilde{J}_{2}V            & \Lambda                     & V^{\dagger}\tilde{J}_{-2}V          & \cdots \\
    \cdots & V^{\dagger}\tilde{J}_{4}V            & V^{\dagger}\tilde{J}_{2}V                   & \Lambda -2 \omega\mathbf{I} & \cdots \\
                  & \vdots                           & \vdots                               & \vdots               & \ddots
\end{bmatrix}.
\end{equation}
We just claim the key difference with the previous one. First, the original basis change from $\{\ket{k=0},\ket{k=\pi}\}$ to two linear combinations of the two vectors$\{\ket{\phi_{1}},\ket{\phi_{2}}\}$. Then, the scattering is not just between $\ket{\phi_{1}},\ket{\phi_{2}}$. There are still couplings between $\ket{\phi_{1}}$ and $e^{-in\omega t}\ket{\phi_{1}}$. However, the series equality can still be applied in this perturbative problem with more diverse physics.
\subsection{Generalization to the accordion-type space-time symmetry}
In this part, we give a detailed derivation of the generalized formula about Hamiltonians with the space-time symmetry
\begin{equation}
    H(x,t) = H(x+\frac{1}{af(t)},t),H(x,t)=H(x,t+T),f(t)=f(t+T).
\end{equation}
There is discrete spatial translational symmetry at any time t, so we can get a Fourier expansion of $x$,
\begin{equation}
    H(x,t)=\sum_{k_{n}}\tilde{H}(k_{n}(t),t)e^{ik_{n}(t)x},k_{n}(t)=2\pi anf(t).
\end{equation}
This is also a periodicity to time, so we calculate the mth-order Fourier coefficients to time,
\begin{equation}
\begin{split}
    \tilde{H}_{m} &=\frac{1}{T}\int_{0}^{T}H(x,t)e^{-im\omega t}dt=\sum_{k_{n}}\frac{1}{T}\int_{0}^{T}\tilde{H}(k_{n}(t),t)e^{i(k_{n}(t)x-m\omega t)}dt\\
    &=\sum_{k_{n},b}\frac{1}{T}\int_{0}^{T}\tilde{H}_{b}(k_{n})e^{ib\omega t}e^{i(k_{n}(t)x-m\omega t)}dt.
    \end{split}
\end{equation}
Finally, we can get the scattering structure in frequency $m\omega$,
\begin{equation}
\begin{split}
    \tilde{H}^{\star}_{m}(q)&\propto \sum_{k_{n},b}\frac{1}{T}\int_{0}^{T}\tilde{H}_{b}(k_{n})\delta(q-k_{n}(t))e^{i(b-m)\omega t}dt\\
    &\propto \sum_{k_{n\neq0},b}\frac{1}{T}\tilde{H}_{b}(k_{n})\frac{e^{i(b-m)\omega f^{-1}(q/2a\pi n)}(f^{-1})'(\frac{q}{2a\pi n})}{\lvert 2a\pi n\rvert}+\tilde{H}_{m}(k=0)\delta(q) .
\end{split}
\end{equation}
This is the formula presented in the paper. The important point here is that $k_{n}(t)=2\pi anf(t)$ is also a periodic function with frequency $\omega$, so we can make a Fourier transformation again.
\begin{figure}[htbp]
    \centering
    \includegraphics[width=0.3\linewidth]{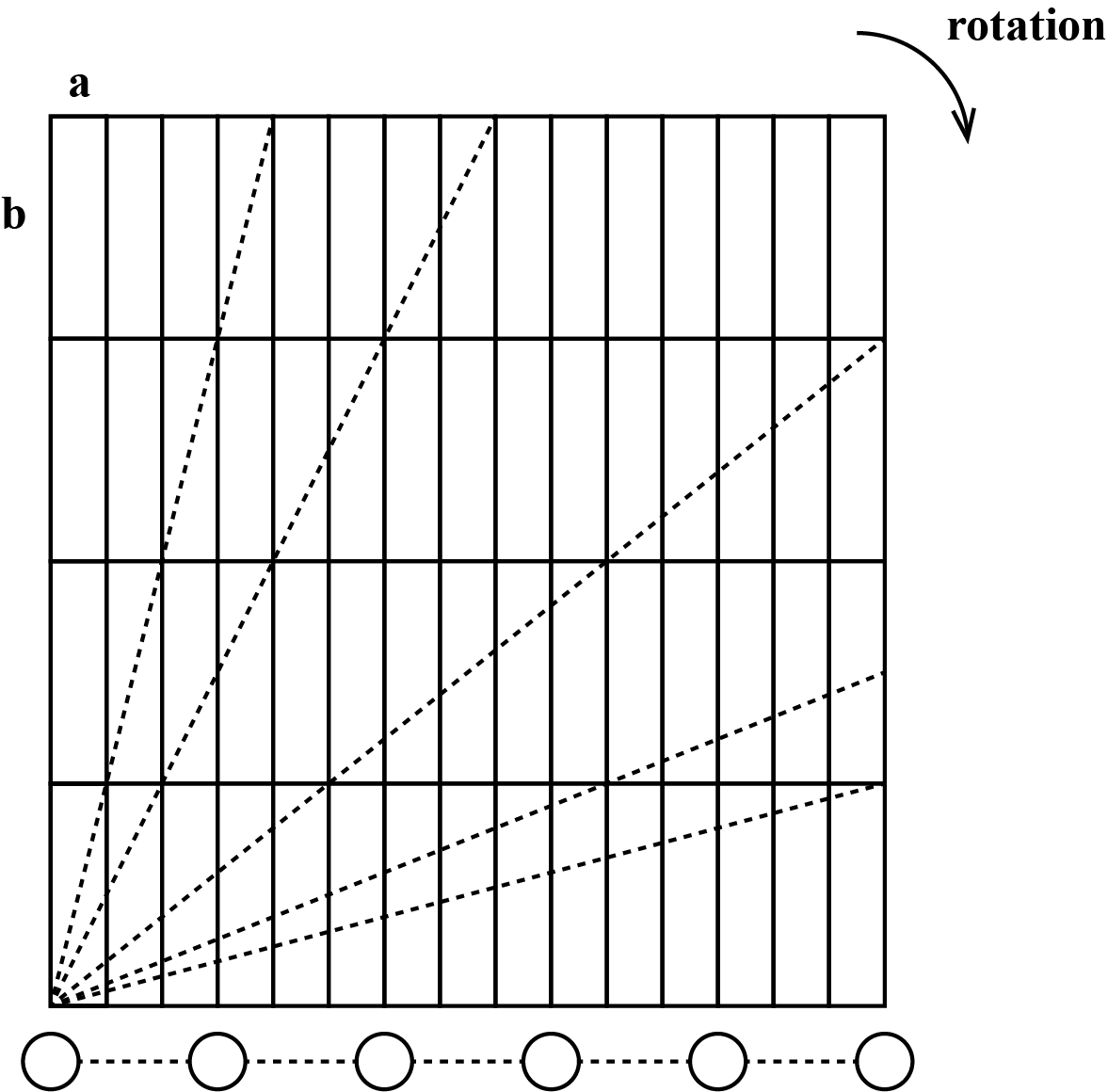}
    \caption{ The illustration of the experimental setup of the accordion-type driving system.}
    \label{fig-experiment}
\end{figure}

\section{Possible experimental realization}
Designing an on-site potential whose spatial periodicity is dependent on time may be a difficult task. Here, we provide a possible experimental realization based on a similar method in the Moiré superlattice. The setup requires a one-dimensional tight-binding lattice and a two-dimensional soft on-site potential as shown in Fig.~\ref{fig-experiment}. Their coupling can only occur when the well of the on-site potential is on the line. When the two-dimensional on-site potential is rotated, the spatial periodicity of it on the lattice will change from $a$ to $\sqrt{a^{2}+b^{2}}$, which behaves as a modulation of spatial periodicity. Although there is incommensurate phenomenon at some angles, they may not influence the major behavior of this model as shown in our numerical results.

\bibliography{references}